\documentclass[nofootinbib,aps,11pt]{revtex4-1}
\usepackage{graphicx}
\usepackage{hyperref}
\usepackage{cancel}
\usepackage{amssymb}
\usepackage{textcomp}
\usepackage{amsmath}
\usepackage{bm}
\usepackage{times}
\usepackage{epsfig}
\usepackage{color}

\newcommand{\tr}{{\rm Tr}}
\newcommand{\eV}{{\rm eV}}
\newcommand{\br}{{\rm Br}}

\newcommand{\GeV}{{\rm GeV}}
\newcommand{\TeV}{{\rm TeV}}

\newcommand{\DM}{{\rm DM}}

\begin{document}
\title{\LARGE Phenomenology of Colored Radiative Neutrino Mass Model and Its Implications on the Cosmic-ray Observations}
\bigskip
\author{Ran Ding~$^{1}$}
\email{dingran@mail.nankai.edu.cn}
\author{Zhi-Long Han~$^{2}$}
\email{sps\_hanzl@ujn.edu.cn}
\author{Li Huang~$^{3,4}$}
\email{huangli@itp.ac.cn}
\author{Yi Liao~$^{5,1,3}$}
\email{liaoy@nankai.edu.cn}
\affiliation{
$^1$ Center for High Energy Physics, Peking University, Beijing 100871, China
\\
$^2$ School of Physics and Technology, University of Jinan, Jinan, Shandong 250022, China
\\
$^3$
Key Laboratory of Theoretical Physics, Institute of Theoretical Physics, Chinese Academy of Sciences, Beijing 100190, China
\\
$^4$ University of Chinese Academy of Sciences, No.19(A) Yuquan Road, Beijing 100049, China
\\
$^5$ School of Physics, Nankai University, Tianjin 300071, China}
\date{\today}

\begin{abstract}
  We extend the colored Zee-Babu model with a gauged $U(1)_{B-L}$ symmetry and a scalar singlet dark matter (DM) candidate $S$. The spontaneous breaking of $U(1)_{B-L}$ leaves a residual $Z_2$ symmetry that stabilizes the DM and generates tiny neutrino mass at the two-loop level with the color seesaw mechanism. After investigating dark matter and flavor phenomenology of this model systematically, we further focus on its imprint on two of cosmic-ray anomalies: the Fermi-LAT gamma-ray excess at the Galactic Center (GCE) and the PeV ultra-high energy (UHE) neutrino events at the IceCube. We found that the Fermi-LAT GCE spectrum can be well fitted by DM annihilation into a pair of on-shell singlet Higgs mediators while being compatible with the constraints from relic density, direct detections as well as dwarf spheroidal galaxies in the Milky Way. Although the UHE neutrino events at the IceCube could be accounted for by resonance production of a TeV-scale leptoquark, the relevant Yukawa couplings have been severely limited by current low energy flavor experiments. We then derive the IceCube limits on the Yukawa couplings by employing its latest 6-year data.

\end{abstract}

\maketitle

\section{Introduction}

The existence of dark matter (DM) and tiny neutrino mass poses an outstanding challenge to both theoretical and experimental particle physics. Although current searches coming from the Large Hadron Collider (LHC) and DM direct detections have imposed stringent limits, their null results have not yet provided powerful guidance to physics beyond the standard model (SM). On the other hand, observations from high energy cosmic rays (CR) may offer another angle to face the challenge. In this paper, we will focus on two of them, i.e., the Fermi-LAT gamma-ray excess at the Galactic Center (GCE) and the PeV ultra-high energy (UHE) neutrino events at the IceCube. We will attempt to interpret the two observations in a colored seesaw extension of the SM which generates radiative neutrino mass and has a cold DM particle built in. But before we embark on that, let us briefly review the current status of the two observations.

The GCE was first reported in Ref.~\cite{Goodenough:2009gk} through analysing the Fermi-LAT data, and the signal significance was confirmed by subsequent analyses~\cite{Hooper:2010mq,Hooper:2011ti,
Abazajian:2012pn,Gordon:2013vta,Abazajian:2014fta,Daylan:2014rsa,
Calore:2014xka}. While astrophysical interpretations like millisecond pulsars or unresolved gamma-ray point sources~\cite{Gordon:2013vta,
Abazajian:2014fta,Yuan:2014rca,Bartels:2015aea,Lee:2015fea} are plausible, DM annihilation remains one of popular interpretations because its thermally averaged cross section and morphology of density distribution match the standard WIMP scenario. In particular, Ref.~\cite{Calore:2014xka} gives a comprehensive and systematic analysis with multiple Galactic gamma ray diffuse emission (GDE) models. Very recently, the Fermi-LAT Collaboration has released their updated analysis~\cite{TheFermi-LAT:2017vmf,Fermi-LAT:2017yoi} and concluded that GCE can be caused by an unresolved pulsar-like sources located in the Galactic bulge which they referred to as Galactic bulge population, while the dark matter interpretation is disfavored since its distribution is not consistent with the morphology detected in their analysis. However, a large population of pulsars should be accompanied with a large population of low-mass X-ray binaries in the same region, which turns out to restrict their contribution only up to $4-23\%$ of the observed gamma-ray excess~\cite{Haggard:2017lyq}. Moreover, analyses of spatial distribution and luminosity function of those sources were inconclusive about the presence of such Galactic bulge population~\cite{Bartels:2017xba}. Therefore, dark matter interpretation of GCE is still competitive.

When using model independent fitting with DM directly annihilated into a pair of SM particles, the GCE spectrum is best fit by the $b\bar{b}$ final state~\cite{Daylan:2014rsa}. The other final states ($\tau^+\tau^-$, $q\bar{q}$ $c\bar{c}$, $gg$, $W^+W^-$, $ZZ$, $hh$ and $t\bar{t}$) with different DM mass and annihilation cross section are also acceptable~\cite{Calore:2014nla,
Agrawal:2014oha,Cline:2015qha,Elor:2015tva}. Additionally, when taking into account uncertainties in DM halo profiles and propagation models, the annihilation cross section required by GCE is compatible with the limits from other indirect DM searches like dwarf spheroidal galaxies (dSphs) of the Milky Way and the antiproton and CMB observations~\cite{Calore:2014nla,Cirelli:2014lwa,
Ade:2015xua,Slatyer:2015jla,Dutta:2015ysa}. The DM annihilation explanation of the GCE has attracted great interest in the past few years and has been extensively explored in various new physics models~\cite{Berlin:2014tja,Alves:2014yha,Agrawal:2014una,Abdullah:2014lla,
Martin:2014sxa,Berlin:2014pya,Basak:2014sza,Cline:2014dwa,Cheung:2014lqa,Ko:2014loa,Cahill-Rowley:2014ora,
Freytsis:2014sua,Kaplinghat:2015gha,Chen:2015nea,Gherghetta:2015ysa,Dutta:2015ysa,Cao:2015loa,Freese:2015ysa,
Duerr:2015bea,Cai:2015tam,Tang:2015coo,Ding:2016wbd,Krauss:2016cdi,Escudero:2017yia}. These models can be classified into two  scenarios from annihilation patterns:
\begin{itemize}
  \item {DM annihilates directly into SM final states,}
  \item {DM annihilates into some intermediate particles, which subsequently cascade decay into SM particles.}
\end{itemize}
While the first scenario usually suffers from stringent constraints from DM direct detections and collider searches, the second has the advantage that cascade decays can soften and broaden the resulting photon spectrum, thus considerably enlarging the parameter space and relaxing the experimental constraints. More interestingly, GCE can also be interpreted in DM models with a global or local $\mathbb{Z}_3$ symmetry by invoking semi-annihilation channels~\cite{Ko:2014loa,Cai:2015tam,Ding:2016wbd}.

The IceCube observatory is a neutrino telescope located at the South Pole, and holds the unique window to cosmic UHE neutrinos. In the 4-year data set released in year 2015, a total of 54 UHE neutrino events are collected (including 39 cascade events and 14 muon track events) with $7\sigma$ excess over the expected atmospheric background~\cite{Aartsen:2015zva}. Particularly, three events with an energy above PeV present a bit of excess on the SM prediction~\cite{Aartsen:2013bka,Aartsen:2013jdh,Aartsen:2014gkd}. Very recently, the IceCube Collaboration has published the preliminary 6-year result~\cite{Aartsen:2017mau}, with the total number of events increased to 82 with 28 of them being observed in the recent two years. Note that all of new events have energies below 200 TeV, and the excess in the PeV range still exists. The origin of these PeV UHE neutrino events remains mysterious and immediately causes great interest in both astrophysics and particle physics communities. While the astrophysics community focuses on various astrophysical sources~\cite{Cholis:2012kq,Anchordoqui:2013dnh,Murase:2014tsa}, the particle physics community tries to relate them to new physics phenomena. For instance, in the models of decaying superheavy DM~\cite{Feldstein:2013kka,Esmaili:2013gha,Ema:2013nda,Bhattacharya:2014vwa,Higaki:2014dwa,
Rott:2014kfa,Esmaili:2014rma,Fong:2014bsa,Murase:2015gea,Aisati:2015vma,Boucenna:2015tra,Ko:2015nma,
Fiorentin:2016avj,Dev:2016qbd,Chianese:2016smc,Cohen:2016uyg,Dhuria:2017ihq}~\footnote{Models of DM annihilation are challenged by the unitarity bound~\cite{Bai:2013nga,Griest:1989wd,Hui:2001wy}.}, a DM particle of PeV mass is required in order to reproduce the desired UHE neutrino events. Such superheavy particles are very difficult to probe in other experiments and thus phenomenologically less interesting. Another possible explanation invokes a new particle resonance in the TeV region~\cite{Doncheski:1997it,Anchordoqui:2006wc,Alikhanov:2013fda,Barger:2013pla,Dutta:2015dka,
Dey:2015eaa,Mileo:2016zeo,Dev:2016uxj}, in accord with the common belief that new physics should appear there. This latter scenario appears phenomenologically advantageous and could be examined with other means, in particular by direct searches at the LHC.

The six orders of magnitude difference in the energy scale between the GCE (GeV) and IceCube (PeV) events makes it challenging to explain them in a single framework. Here we present a novel example for this issue. We extend the colored Zee-Babu model~\cite{Babu:2001ex} with a $U(1)_{B-L}$ gauge symmetry and a singlet scalar DM candidate. Another singlet Higgs scalar associated with the $U(1)_{B-L}$ symmetry serves as an on-shell mediator for DM annihilation resulting in the GCE spectrum, while the leptoquark (LQ) is responsible for the resonance production of extra UHE neutrino events. The same singlet Higgs scalar and leptoquark generates tiny neutrino mass at two loops. In the next section we describe the model and discuss relevant experimental constraints on its parameter space. Sections~\ref{sec:DM&GCE} and~\ref{sec:IceCube} include the core contents of this work, in which the DM properties, GCE spectrum and UHE neutrino event rate at IceCube are systematically investigated. In section~\ref{sec:DM&GCE}, we explore the vast parameter space that satisfies the constraints from relic abundance and direct detections, and discuss the dominant annihilation channels. A comprehensive fit to the GCE spectrum is then presented incorporating all these limits. In section~\ref{sec:cs}, we calculate the SM and LQ contributions to the neutrino-nucleon scattering cross section. Then in section~\ref{sec:event}, we estimate the LQ contribution to the UHE neutrino event rate at IceCube and perform a likelihood analysis to determine the parameter space. Finally, we draw our conclusion in section~\ref{sec:con}.

\section{Model and Relevant Constraints}\label{sec:model}

\subsection{The Model}

The particle contents and their charge assignments are shown in Table. \ref{Tab:Content}. In addition to the LQ $\psi$ and diquark $\omega$, we further introduce two singlet scalars, $\varphi$ with lepton number $L=2$ and $S$ with $L=\frac{1}{2}$. Here, $\varphi$ is used to break the $U(1)_{B-L}$ gauge symmetry spontaneously, thus generating the $L$-breaking trilinear term $\psi^*\psi^*\omega$ required for radiative neutrino masses. Notably, due to the proper charge assignment of $S$, the $U(1)_{B-L}$ symmetry forbids any gauge invariant terms that would allow $S$ to decay, promoting $S$ a DM candidate without imposing ad hoc discrete symmetry~\cite{Rodejohann:2015lca,Biswas:2016ewm,Klasen:2016qux}. In order to make $U(1)_{B-L}$ anomaly free, some fermions neutral under the SM gauge group but with exotic $B-L$ charges other than $-1$ could be employed~\cite{Montero:2007cd,Wang:2015saa,Patra:2016ofq,Wang:2017mcy,
Nanda:2017bmi,Han:2017ars}.

\begin{table}[!htbp]\large
\begin{tabular}{|c|c|c|c|c|c|c||c|c|c|c|}
\hline\hline
 & $Q_{L}$ & $u_R$ & $d_R$ & $L_{L}$ & $\ell_R$ & $\Phi$ & $\psi$ & $\omega$ & $~\varphi ~$ & $~S ~$
\\ \hline\hline
$SU(3)_C$ & $3$ & $3$ & $3$ & $1$ & $1$ & $1$ & $3$ & $6$ & $1$ & $1$
\\ \hline
$SU(2)_L$ & $2$ & $1$ & $1$ & $2$ & $1$ & $2$ & $1$ & $1$ & $1$ & $1$
\\ \hline
$U(1)_Y$ & $\frac{1}{6}$ & $\frac{2}{3}$ & $-\frac{1}{3}$ & $-\frac{1}{2}$ & $-1$ & $\frac{1}{2}$ & $-\frac{1}{3}$ & $-\frac{2}{3}$ & $0$ & $0$
\\ \hline
$U(1)_{B-L}$ & $\frac{1}{3}$ & $\frac{1}{3}$ & $\frac{1}{3}$ & $-1$ & $-1$ & $0$ & $-\frac{2}{3}$  & $\frac{2}{3}$ & $-2$ & $-\frac{1}{2}$
\\ \hline \hline
$L$ & $0$ & $0$ & $0$ & $1$ & $1$ & $0$ & $1$ & $0$ & $2$ & $\frac{1}{2}$
\\ \hline
$B$ & $\frac{1}{3}$ & $\frac{1}{3}$ & $\frac{1}{3}$ & $0$ & $0$ & $0$ & $\frac{1}{3}$ & $\frac{2}{3}$ & $0$ & $0$
\\ \hline\hline
\end{tabular}
\caption{Particle contents and their charge assignments. The double vertical line separates the SM particles from the new ones.}
\label{Tab:Content}
\end{table}

The relevant Yukawa interactions involving the LQ $\psi$ and the diquark $\omega$ are given by
\begin{equation}
-\mathcal{L}_{\text{Y}}= y_L^{ij} \overline{(L_{Li})^C}i\sigma_2 Q_{Lj} \psi^{*} + y_R^{ij} \overline{(\ell_{Ri})^C}u_{Rj} \psi^{*} + y_{\omega}^{ij} \overline{(d_{Ri})^C} d_{Rj} \omega^{*}+ y_{\psi}^{ij} \overline{(u_{Ri})^C} d_{Rj} \psi+\text{h.c.},
\label{eq:yukawa}
\end{equation}
where $\sigma_2$ is the second Pauli matrix, $ij$ refers to the SM generations, and the color indices are suppressed. Here, $y_\omega$ is a symmetric matrix, while $y_{L,R}$ and $y_\psi$ are general complex matrices. The neutrinos interact with the LQ only through the $y_L$ term, which induces neutrino masses at the two-loop level as shown in Fig.~\ref{Fig:mv}. Compared to the original Zee-Babu model, no antisymmetric Yukawa couplings are involved in neutrino mass generation so that all neutrino masses can be non-zero in this colored Zee-babu model. And the $y_\psi$ together with the $y_{L,R}$ terms can lead to the tree-level proton decay~\cite{Arnold:2013cva}. In principle, this $y_\psi$ term can be forbidden by some discrete symmetry~\cite{Chang:2016zll}. For simplicity, we will assume $y_\psi=0$ in the following discussion. Note that due to the charge assignments the two scalar singlets $\varphi$ and $S$ do not couple to fermions at the Lagrangian level.

The gauge invariant scalar potential is described by
\begin{eqnarray}\label{Eq:sp}
V & = & - \mu^2_{\Phi} \Phi^\dag \Phi -\mu^2_{\varphi} \varphi^{\dag} \varphi + \mu_{S}^2 S^{\dag}S
+\mu^2_{\psi} \psi^\dag\psi + \mu^2_{\omega} \tr(\omega^\dag \omega) \\ \nonumber
&& + \lambda_{\Phi} (\Phi^\dag \Phi)^2  + \lambda_{\varphi} (\varphi^{\dag} \varphi)^2 +\lambda_S (S^{\dag} S)^2 +\lambda_\psi (\psi^\dag \psi)^2 + \lambda_{\omega} [\tr(\omega^\dag \omega)]^2
 \\ \nonumber
 && + \lambda_{\Phi \varphi} (\Phi^\dag \Phi) (\varphi^{\dag} \varphi)  + \lambda_{Sh} (\Phi^\dag \Phi) (S^{\dag} S) + \lambda_{\Phi \psi} (\Phi^\dag \Phi)(\psi^\dag \psi)+ \lambda_{\Phi \omega} (\Phi^\dag \Phi)\tr(\omega^\dag \omega) \\ \nonumber
 &&+ \lambda_{S H_0} (\varphi^{\dag} \varphi) (S^{\dag} S)+ \lambda_{\varphi \psi} (\varphi^{\dag} \varphi)(\psi^\dag \psi)+\lambda_{\varphi \omega} (\varphi^{\dag} \varphi)\tr(\omega^\dag \omega)+\lambda_{S \psi} (S^{\dag} S)(\psi^\dag \psi) \\ \nonumber
 &&+\lambda_{S \omega} (S^{\dag} S) \tr(\omega^\dag \omega) + \lambda_{\psi \omega} (\psi^\dag \psi) \tr(\omega^\dag \omega) + [\sqrt{2}\lambda~\varphi \psi^{*} \psi^{*} \omega+\text{h.c.}],
\end{eqnarray}
where $\mu_{X}^2(X=\Phi,\varphi,S,\psi,\omega)$ are all taken to be positive, and the trace is over the color indices. In this way, the $SU(2)_L\times U(1)_Y$ and $U(1)_{B-L}$ gauge symmetries are spontaneously broken by the vacuum expectation values of $\Phi$ and $\varphi$, respectively. Due to the $B-L$ charge assignment of $S$, one can still have $\langle S\rangle=0$ after spontaneous symmetry breaking, so that a residual $Z_2$ symmetry remains under which only $S$ is odd. This blocks all potential decays of $S$, making it a viable DM candidate~\cite{Rodejohann:2015lca,Biswas:2016ewm,Klasen:2016qux}.

In unitary gauge the scalar fields $\Phi$ and $\varphi$ are denoted as
\begin{align}
\Phi=\frac{v_{\phi}+\phi^0}{\sqrt{2}}\left(
\begin{array}{c}
0\\
1
\end{array}\right),
\quad \varphi = \frac{v_{\varphi}+ \varphi^0}{\sqrt{2}}.
\end{align}
Here $v_\phi=246~\GeV$ is the electroweak scale, and the vacuum expectation value (VEV) $v_\varphi$ generates the mass for the new gauge boson $Z'$ of $U(1)_{B-L}$,
\begin{equation}
M_{Z'}=2 g_{BL} v_\varphi,
\end{equation}
where $g_{BL}$ is the gauge coupling of $U(1)_{B-L}$. The LEP bound requires that~\cite{Cacciapaglia:2006pk}
\begin{equation}
M_{Z'}/g_{BL} = 2 v_\varphi \gtrsim 7~\TeV,
\end{equation}
yielding a lower limit on $v_\varphi\gtrsim 3.5~\TeV$. On the other hand, the direct searches for the $Z'$-boson at LHC in the dilepton channel have excluded $M_{Z'}\lesssim 4~\TeV$~\cite{Aaboud:2017buh,ATLAS:2016cyf,CMS:2016abv}, and recasting these searches in the gauged $U(1)_{B-L}$ model has been performed in Refs.~\cite{Okada:2016gsh,Okada:2016tci,DeRomeri:2017oxa} to acquire the exclusion region in the $M_{Z'}-g_{BL}$ plane. Considering these bounds, we choose to work with $M_{Z'}=4~\TeV$ and $g_{BL}=0.1$, so that $v_{\varphi}=20~\TeV$ in our following discussion. The masses of the DM $S$, LQ $\psi$ and diquark $\omega$ can be figured out from the scalar potential in Eq.~(\ref{Eq:sp}):
\begin{eqnarray}
M_{S}^2 & = & \mu_S^2 + \frac{\lambda_{\Phi S}}{2} v_\phi^2+ \frac{\lambda_{\varphi S}}{2} v_\varphi^2,\\
M_{\psi}^2 & = & \mu_\psi^2  + \frac{\lambda_{\Phi \psi}}{2} v_\phi^2 + \frac{\lambda_{\varphi \psi}}{2} v_\varphi^2, \\
M_{\omega}^2 & = & \mu_\omega^2  + \frac{\lambda_{\Phi \omega}}{2} v_\phi^2 + \frac{\lambda_{\varphi \omega}}{2} v_\varphi^2.
\end{eqnarray}
In this work, we will consider $M_S$ in the interval $[5,150]~\GeV$ and $[500,1500]~\GeV$ for the low and high mass region, respectively. The constraints from relic density and direct detections will be discussed in Sec.~\ref{sec:DM&GCE}. Assuming the LQ $\psi$ decaying exclusively into $eq$, $\mu q$, and $\tau q$, CMS (ATLAS) has excluded $M_\psi<1010,~1165,~850~\GeV$~\cite{Khachatryan:2015vaa,CMS:2016qhm,
Khachatryan:2016jqo} ($M_\psi<1100,~1050,~534~\GeV$~\cite{Aaboud:2016qeg,Aad:2011ch,
Aaboud:2016qeg,ATLAS:2012aq,ATLAS:2013oea}). However, both $\psi\to \ell q$ and $\psi\to \nu_\ell q'$ exist in our model. The maximum exclusion limits by CMS (ATLAS) for the first and second generation LQ are $850,~960~\GeV$~\cite{Khachatryan:2015vaa,CMS:2016qhm} ($900,~830~\GeV$~\cite{Aaboud:2016qeg}) when assuming $\text{BR}(\psi\to\ell q)=\text{BR}(\psi\to \nu_\ell q')=0.5$ with $\ell=e$ or $\mu$, respectively. ATLAS has also excluded $M_\psi<625~\GeV$ when BR($\psi\to \nu_\tau b)=1$ for the third generation LQ\cite{Aad:2015caa}. As for the scalar diquark $\omega$, CMS has excluded $M_\omega\lesssim 7~\TeV$~\cite{Khachatryan:2015dcf,Aaboud:2017yvp}. In the following, we will mainly consider $M_{\psi}\gtrsim1~\TeV$ and $M_{\omega}=7~\TeV$ to respect these collider limits.

The $\lambda_{\Phi \varphi}$ term induces mixing between $\phi^0$ and $\varphi^0$, with the squared mass matrix given by
\begin{align}
\mathcal{M}^2_0=\left(
\begin{array}{cc}
2\lambda_\Phi v_\phi^2 & \lambda_{\Phi \varphi} v_\phi v_\varphi\\
\lambda_{\Phi \varphi} v_\phi v_\varphi & 2\lambda_\varphi v_\varphi^2
\end{array}\right),
\end{align}
which is diagonalized to the mass eigenstates $(h,H_0)$
\begin{eqnarray}
h&=& \phi^0 \cos\theta + \varphi^0 \sin\theta, \\
H_0 &=& \varphi^0 \cos\theta-\phi^0 \sin\theta.
\end{eqnarray}
by an angle $\theta$ determined by
\begin{equation}
\tan 2\theta = \frac{\lambda_{\Phi \varphi} v_\phi v_\varphi}
{\lambda_\Phi v_\phi^2-\lambda_\varphi v_\varphi^2},
\end{equation}
with $-\pi/4<\theta<\pi/4$. The masses of $h$, $H_0$ are
\begin{eqnarray}
M_h^2 & = & \lambda_\Phi v_\phi^2 + \lambda_\varphi v_\varphi^2 + (\lambda_\Phi v_\phi^2 - \lambda_\varphi v_\varphi^2)/\cos(2\theta),\\
M_{H_0}^2 & = & \lambda_\Phi v_\phi^2 + \lambda_\varphi v_\varphi^2 -(\lambda_\Phi v_\phi^2 - \lambda_\varphi v_\varphi^2)/\cos(2\theta).
\label{eq:mhiggs}
\end{eqnarray}
Here $h$ is regarded as the Higgs boson with $M_h=125~\GeV$ discovered at LHC~\cite{Aad:2012tfa,Chatrchyan:2012xdj,Aad:2015zhl}. According to previous studies on scalar singlets, in the high mass region $M_{H_0}>500~\GeV$~\cite{Barger:2006sk,Barger:2007im,Robens:2015gla,
Robens:2016xkb}, a small mixing angle $|\sin\theta|\lesssim 0.2$ is allowed by various experimental bounds. In light of the recent Fermi-LAT GCE, we will also consider the low mass region $M_{H_0}\in[5,150]~\GeV$. In this region, the LHC SM Higgs signal rate measurement has excluded $|\sin\theta|\gtrsim 0.36$~\cite{Robens:2016xkb,Giardino:2013bma,Khachatryan:2016vau}, and the LEP search for $ZH_0$ associated production has excluded $|\sin\theta|\gtrsim 0.2$ when $H_0\to b\bar{b}$ dominates~\cite{Abdallah:2004wy}. Thus, it is safe to consider $|\sin\theta|\lesssim 0.1$ in the following discussion.
For convenience, we express the Lagrangian parameters $\lambda_{\Phi,\varphi,\Phi\varphi}$ and $\mu_{\Phi,\varphi}$ in terms of the physical scalar masses $M_{h,H_0}$, mixing angle $\theta$ as well as the VEVs $v_{\phi,\varphi}$:
\begin{eqnarray}
\lambda_{\Phi\varphi} &=& \frac{1}{v_\phi v_\varphi}
(M_h^2-M_{H_0}^2)\cos\theta\sin\theta,\\
\lambda_{\Phi} &=& \frac{1}{4v_\phi^2}
\left[M_h^2+M_{H_0}^2+(M_h^2-M_{H_0}^2)\cos2\theta\right], \\
\lambda_\varphi &=& \frac{1}{4v_\varphi^2}
\left[M_h^2+M_{H_0}^2+(M_{H_0}^2-M_h^2)\cos2\theta\right],\\
\mu_\phi^2 & = & \frac{1}{4 v_\phi}
\left[(M_h^2+M_{H_0}^2)v_\phi + (M_h^2-M_{H_0}^2) (v_\phi \cos2\theta+v_\varphi\sin2\theta)\right],\\
\mu_\varphi^2 & = & \frac{1}{4 v_\varphi}
\left[(M_h^2+M_{H_0}^2)v_\varphi + (M_{H_0}^2-M_h^2) (v_\varphi \cos2\theta-v_\phi\sin2\theta)\right].
\end{eqnarray}

\subsection{Neutrino Mass}

\begin{figure}[!htbp]
\begin{center}
\includegraphics[width=0.5\linewidth]{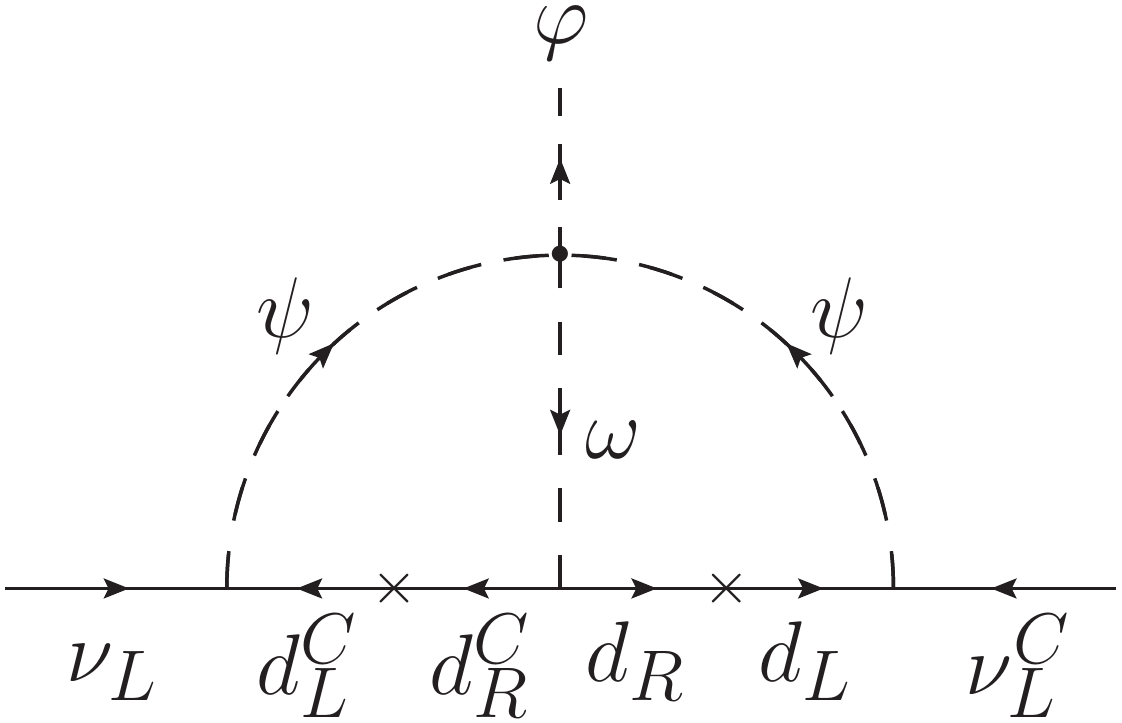}
\end{center}
\caption{Two-loop generation of neutrino mass.
\label{Fig:mv}}
\end{figure}

As shown in FIG.~\ref{Fig:mv}, the neutrino masses are induced at two loops \cite{Chang:2016zll}:
\begin{equation}\label{Eq:mv}
m_\nu^{ij} = 24 \lambda v_\varphi y_L^{im} M_{d_m} I_{mn} (y_\omega^\dag)^{mn} M_{d_n} (y_L^T)^{nj},
\end{equation}
where the full analytical form for the loop function $I_{mn}$ can be found in Ref.~\cite{McDonald:2003zj}. Considering that the down-type quarks are much lighter than the colored scalars, it can be simplified for order of magnitude estimate to
\begin{equation}
I_{mn} = \frac{1}{(16\pi^2)^2} \frac{1}{M_\omega^2} \frac{\pi^2}{3} I\left(\frac{M_\omega^2}{M_\psi^2}\right),
\end{equation}
where
\begin{eqnarray}
I(x)=\left\{
\begin{array}{lcc}
1+\frac{3}{\pi^2}(\ln^2 x -1)&\text{for}&x\gg1\\
1 &\text{for}&x\to 0
\end{array} \right. .
\end{eqnarray}
Typically, a neutrino mass $m_\nu\sim 0.01~\eV$ can be realised with $\lambda\sim0.1$, $y_L\sim y_\omega\sim0.01$ when $v_\varphi=20~\TeV$, $M_b=4.7~\GeV$, $M_\psi=1~\TeV$, and $M_\omega=7~\TeV$.
The radiative correction to the masses $M_{\psi}$ and $M_\omega$ involves also the trilinear coupling $\lambda v_\varphi \psi^*\psi^*\omega$, the choice of $\lambda\sim 0.1$ and  $v_\varphi=20~\TeV$ also satisfies the perturbativity requirement $\lambda v_\varphi\lesssim5\min(M_\psi,M_\omega)$ for $M_\psi\sim 1~\TeV$ and $M_\omega\sim 7~\TeV$~\cite{Babu:2002uu,Nebot:2007bc}. The neutrino mass in Eq.~(\ref{Eq:mv}) can be written in a compact form
\begin{equation}
m_\nu = y_L \Omega y_L^T,
\end{equation}
where $\Omega^{mn}=\lambda v_\varphi M_{d_m} (y_\omega^\dag)^{mn} M_{d_n} I(M_{\omega}^2/M_\psi^2)/(32\pi^2 M^2_\omega)$. In principle, by adopting a proper parametrization~\cite{Casas:2001sr,Liao:2009fm}, the Yukawa coupling $y_L$ can be solved in terms of the neutrino masses, mixing angles and a generalized orthogonal matrix with three free parameters, so that the neutrino oscillation data can be automatically incorporated. Following this approach, a benchmark point has been suggested in Ref.~\cite{Kohda:2012sr}; see Ref.~\cite{Chang:2016zll} for more details. As to be discussed below, in this work we follow the usual phenomenological practice to take Yukawa components $y_L^{ij}$ as input parameters whose values will be constrained by IceCube data and low-energy experiments.

\subsection{Flavor Constraints}
\label{sec:flavor}

The LQ $\psi$ can induce various flavor violating processes at the tree level. To minimize such processes, one usually assumes $y_R=0$~\cite{Kohda:2012sr,Chang:2016zll}, since the $y_R$ term is less important to neutrino masses as well. This also fits our interest in the IceCube UHE neutrino events which may be induced by $y_L$ but not $y_R$ couplings. Since LQ is heavy, its effects can be incorporated into effective four-fermion operators of the SM leptons and quarks. The constraints on these operators have been studied in Ref.~\cite{Carpentier:2010ue} for the normalized Wilson coefficients:
\begin{equation}
\epsilon_{ijkn} = \frac{y_L^{ik}y_L^{jn}}{4\sqrt{2} G_F M_\psi^2}.
\end{equation}
The relevant upper limits on $\epsilon_{ijkn}$ in the colored Zee-Babu model are summarized in Table 3 of Ref.~\cite{Chang:2016zll}. In particular, there are two $\epsilon_{ijkn}$ that are strongly constrained: one is $\epsilon_{e\mu uu}<8.5\times10^{-7}$ from $\mu$-$e$ conversion in nuclei, and the other is $\epsilon_{\ell\ell^\prime\! uc}<9.4\times10^{-6}$ from the $K$-meson decay. This indicates that \cite{Nomura:2016ask}
\begin{eqnarray}
y_L^{eu} y_L^{\mu u} = 4\sqrt{2}G_F M_\psi^2\, \epsilon_{e\mu uu} < 5.6\times10^{-5}\left( \frac{M_\psi}{1~\TeV}\right)^2, \\
y_L^{\ell u} y_L^{\ell^\prime\! c} = 4\sqrt{2}G_F M_\psi^2\, \epsilon_{\ell\ell^\prime uc} < 6.2\times10^{-4}\left( \frac{M_\psi}{1~\TeV}\right)^2.
\end{eqnarray}
One way to satisfy these bounds is to assume, e.g., $y_L^{\ell u}\lesssim 0.001$ and $y_L^{\ell c}\lesssim0.1$ at $M_\psi\sim1~\TeV$. The constraints on other components of $\epsilon_{ijkn}$ are quite loose, and can be readily avoided by, e.g., $y_L^{\ell q}\lesssim \mathcal{O}(0.1)$ for a TeV scale $M_\psi$~\cite{Guo:2017gxp}.

The Yukawa coupling $y_L^{ij} \overline{(L_{Li})^C}i\sigma_2 Q_{Lj} \psi^{*}$ is also responsible for lepton flavor violation (LFV) processes at one loop. According to Ref.~\cite{Chang:2016zll}, the constraints from the radiative decay $\ell\to \ell^\prime \gamma$ are usually more stringent than other LFV processes, and the branching ratio is calculated as~\cite{Chang:2016zll,Dorsner:2016wpm}
\begin{equation}\label{LFV}
\text{BR}(\ell\to\ell^\prime \gamma)=\text{BR}(\ell\to \ell^\prime \bar{\nu}_{\ell^\prime} \nu_\ell)
 \frac{3\alpha N_C^2}{16\pi G_F^2 M_\psi^4}\frac{|A_L^{\ell \ell^\prime}|^2+|A_R^{\ell \ell^\prime}|^2}{M_{\ell}^2},
\end{equation}
where $N_C=3$ and the LQ-quark loop yields
\begin{eqnarray}\label{AR}
A_R^{\ell \ell^\prime}= - \sum_{q=u,c,t} \Big[ \left(y_L^{\ell^\prime\! q*} y_L^{\ell q} M_{\ell} + y_R^{\ell^\prime\! q*}y_R^{\ell q} M_{\ell^\prime}\right)F_1(r_q) +y_L^{\ell^\prime\! q*} y_R^{\ell q} M_q F_2(r_q)\Big].
\end{eqnarray}
Here $r_q=M_q^2/M_\psi^2$, $A_L^{\ell \ell^\prime}=A_R^{\ell \ell^\prime}|_{y_L\leftrightarrow y_R}$, and the loop functions are \cite{Chang:2016zll}
\begin{eqnarray}
F_1(x)&=& \frac{1}{12(1-x)^4}\left[1+4x-5x^2+2x(2+x)\ln x\right],\\
F_2(x)&=&\frac{1}{6(1-x)^3}\left[7-8x+x^2+2(2+x)\ln x\right].
\end{eqnarray}
In the limit $x\to 0$, the loop functions behave as $F_1(x)\to 1/12$ and $F_2(x)\to (7+4\ln x)/6<0$. If $y_L^{\ell q}\sim y_R^{\ell q}$, the second term in Eq.~(\ref{AR}) is expected to be dominant, since $|M_\ell F_1(r_q)|\ll|M_q F_2(r_q)|$. Hence we assume $y_R=0$ in numerical analysis partly for minimizing the LQ contribution to lepton radiative decays. With this assumption, $A_R^{\ell\ell^\prime}$ dominates over $A_L^{\ell\ell^\prime}$ considering $M_\ell \gg M_{\ell^\prime}$, and Eq.~(\ref{LFV}) simplifies to
\begin{equation}
\text{BR}(\ell\to\ell^\prime \gamma)=\text{BR}(\ell\to \ell^\prime \bar{\nu}_{\ell^\prime} \nu_\ell)
 \frac{3\alpha N_C^2}{16\pi G_F^2 M_\psi^4}\left|\sum_{q=u,c,t}y_L^{\ell^\prime\! q*} y_L^{\ell q}F_1(r_q)\right|^2.
\end{equation}

Currently, the most stringent limits on lepton radiative decays are $\text{BR}(\mu\to e\gamma)<4.2\times10^{-13}$~\cite{TheMEG:2016wtm}, $\text{BR}(\tau\to \mu\gamma)<4.4\times10^{-8}$~\cite{Aubert:2009ag}, and $\text{BR}(\tau\to e\gamma)<3.3\times 10^{-8}$~\cite{Aubert:2009ag}. They translate into the constraints on the Yukawa couplings
\begin{eqnarray}
\left|\sum_{q=u,c,t} y_L^{eq*} y_L^{\mu q}  \right|&\lesssim&1.4 \times 10^{-3} \left(\frac{M_\psi}{1~\TeV}\right)^2,\\
\left|\sum_{q=u,c,t} y_L^{\mu q *} y_L^{\tau q} \right|&\lesssim&  1.1 \left(\frac{M_\psi}{1~\TeV}\right)^2,\\
\left|\sum_{q=u,c,t} y_L^{eq*} y_L^{\tau q}\right|&\lesssim& 0.98 \left(\frac{M_\psi}{1~\TeV}\right)^2.
\end{eqnarray}
For a flavor universal structure, the above requires $|y_L^{\ell q}|\lesssim 0.02$ at $M_\psi\sim 1~\TeV$. On the other hand, a hierarchal structure $|y_L^{eq}|\ll|y_L^{\mu q}|\sim|y_L^{\tau q}|\sim\mathcal{O}(0.1)$ is still allowed at $M_\psi\sim 1~\TeV$, because radiative $\tau$ decays are less stringently constrained~\cite{Guo:2017gxp}.

A by-product of lepton radiative decays is the LQ contribution to the anomalous magnetic moment of the charged lepton $\ell$~\cite{Chakraverty:2001yg,Cheung:2001ip}
\begin{equation}
\Delta a_{\ell} = -\frac{ N_C M_{\ell}}{ 8\pi^2 M_\psi^2} \sum_{q=u,c,t}
\left[M_{\ell}\left(|y_L^{\ell q}|^2+|y_R^{\ell q}|^2\right)F_1(r_q)+ M_q \text{Re}(y_L^{\ell q*}y_R^{\ell q}) F_2(r_q)\right].
\end{equation}
Under constraints from LFV, the predicted values are $\Delta a_e=-2\times 10^{-19}$, $\Delta a_\mu=-1\times 10^{-14}$, and $\Delta a_\tau=-2\times 10^{-12}$ for universal Yukawa couplings $|y_L^{\ell q}|\sim 0.01$ at $M_\psi\sim 1~\TeV$ and assuming $y_R=0$, which are far below the current experimental limits~\cite{Giudice:2012ms,Bennett:2006fi}. It is also clear that with the assumption of $y_R=0$ the observed discrepancy $\Delta a_\mu=(27.8\pm 8.8)\times 10^{-10}$~\cite{Bennett:2006fi} cannot be explained, since the contribution of the $|y_L^{\ell q}|^2$ term is negative. To resolve the discrepancy, a nonzero $y_R$ is necessary, e.g., with $y_R^{\mu c}\sim y_R^{\mu t}\sim 0.01$, $y_L^{\mu c}\sim 2.4$, $y_L^{\mu t}\sim0.5$, and $M_\psi\sim 1~\TeV$ \cite{Bauer:2015knc}.

If $\text{Im}(y_L^{\ell q*}y_R^{\ell q})$ is nonzero, the LQ also contributes to the electric dipole moment (EDM) of the charged lepton $\ell$ at one loop~\cite{Cheung:2001ip}
\begin{equation}
d_\ell = \frac{e N_C}{16\pi^2 M_\psi^2} \sum_{q=u,c,t} M_q \text{Im}(y_L^{\ell q*} y_R^{\ell q}) F_2(r_q).
\end{equation}
Typically for $|y_L^{eq}|\sim|y_R^{eq}|\sim 0.01$, $M_\psi\sim 1~\TeV$, and an order one CP phase, the top quark would dominate and contribute to the electron EDM $|d_e|\sim10^{-24} e$-cm, which has already been excluded by the current limit $|d_e|<8.7\times 10^{-29} e$-cm~\cite{Baron:2013eja}. If we still assume $y_R=0$, the EDM will arise at three loops, whose order of magnitude is~\cite{Chang:2016zll}
\begin{equation}
d_\ell \sim \frac{e \alpha N_C}{(16\pi)^3}\frac{M_\ell}{M_\psi^2} \text{Im}
\left[y_L^{\ell k} V_{\text{CKM}}^{kj} (y_L^\dag)^{ji} U_{\text{PMNS}}^{i\ell}\right].
\end{equation}
For $|y_L^{eq}|\sim 0.01$, $M_\psi\sim 1~\TeV$, and an order one combined CP phase, one has $|d_e|\sim 10^{-37}e$-cm, which is much smaller than the current limit.

As for the diquark $\omega$,  the Yukawa couplings $y_{\omega}^{ij}$ are tightly constrained by neutral mesons mixings~\cite{Bona:2007vi}. The corresponding Wilson coefficients for the $K^0$-$\overline{K^0}$, $B_d^0$-$\overline{B_d^0}$ and $B_s^0$-$\overline{B_s^0}$ mixings are respectively
\begin{eqnarray}
\tilde{C}_K^1 &=& - \frac{1}{2M_\omega^2} y_\omega^{11} y_\omega^{22*},\\
\tilde{C}_{B_d}^1 &=& - \frac{1}{2M_\omega^2} y_\omega^{11} y_\omega^{33*},\\
\tilde{C}_{B_s}^1 &=& - \frac{1}{2M_\omega^2} y_\omega^{22} y_\omega^{33*}.
\end{eqnarray}
The 95\% C.L. limits, $|\tilde{C}_K^1|<9.6\times 10^{-13}$, $|\tilde{C}_{B_d}^1|<2.3\times 10^{-11}$, and $|\tilde{C}_{B_s}^1|<1.1\times 10^{-9}$ in units of $\GeV^{-2}$~\cite{Carpentier:2010ue}, then require
\begin{eqnarray}
|y_\omega^{11}y_\omega^{22*}|<1.9\times10^{-6} \left(\frac{M_\omega}{1~\TeV}\right)^2,\\
|y_\omega^{11}y_\omega^{33*}|<4.6\times10^{-5} \left(\frac{M_\omega}{1~\TeV}\right)^2,\\
|y_\omega^{22}y_\omega^{33*}|<2.2\times10^{-3} \left(\frac{M_\omega}{1~\TeV}\right)^2.
\end{eqnarray}
With $M_\omega=7~\TeV$, such constraints correspond to $y_\omega^{ij}\lesssim 0.009$ for a universal Yukawa structure.

\section{DM phenomenology and GCE spectrum fitting}\label{sec:DM&GCE}

\begin{table*}[hbtp]
\begin{tabular}{|c|c|c|c|c|c|c|c|c|c|}
\hline
&$M_S$ & $M_{H_0}$ & $M_{Z^\prime}$ & $g_{BL}$ & $|\theta|$ & $\lambda_{Sh}$ & $\lambda_{SH_0}$ & $\lambda_{S\psi}$ & $M_\psi$\\\hline
Low mass DM & $[5,150]$ & $[5,150]$ & $4000$ & $0.1$ & $[10^{-3},0.1]$ & $[10^{-4},0.1]$ & $[10^{-4},0.1]$ & $0.5$ & $1000$ \\\hline
High mass DM& $[500,1500]$ & $50$ & $4000$ & $[10^{-2},0.5]$ & $10^{-3}$ & $[10^{-3},0.5]$ & $10^{-3}$ & $[10^{-2},0.5]$ & $[500,1500]$ \\\hline
\end{tabular}
\caption{The ranges or values of the input parameters used in DM scan. All masses in units of GeV and $M_h=125~\GeV$.}
\label{tab:DMscan}
\end{table*}

In order to investigate DM phenomenology, we use {\tt FeynRules}~\cite{feynrules} to generate the {\tt CalcHEP}~\cite{Belyaev:2012qa} model file and implement it into the {\tt micrOMEGAs4.3.2} package~\cite{Belanger:2014vza} to calculate the DM relic abundance and DM-nucleon scattering cross section. We perform random scan for parameter space in both low and high mass DM scenarios (with $3\times 10^5$ samples for each), with input parameters shown in Table~\ref{tab:DMscan}. The constraints from DM relic abundance and direct detection experiments are imposed on each sample. For DM relic abundance, we adopt the combined Planck+WP+highL+BAO result in the $2\sigma$ range, $0.1153<\Omega_\DM h^2<0.1221$~\cite{Ade:2013zuv}. For direct detections, we use the latest spin-independent limits obtained by LUX~\cite{Akerib:2016vxi}, XENON1T~\cite{Aprile:2017iyp} and PandaXII~\cite{Cui:2017nnn} Collaborations.

\begin{figure}
\begin{center}
\includegraphics[width=0.88\linewidth]{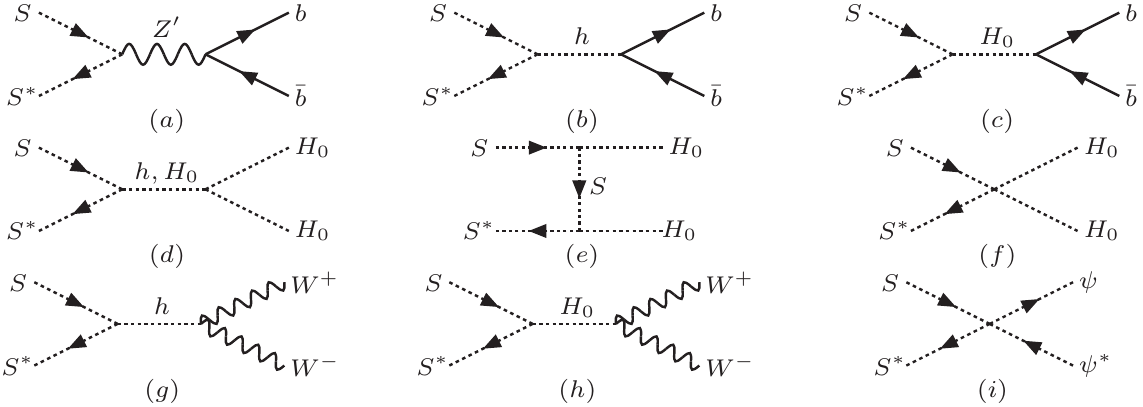}
\end{center}
\caption{Feynman diagrams for various annihilation processes.}
\label{fig:channels}
\end{figure}

\begin{figure}
\begin{center}
\includegraphics[width=0.48\linewidth]{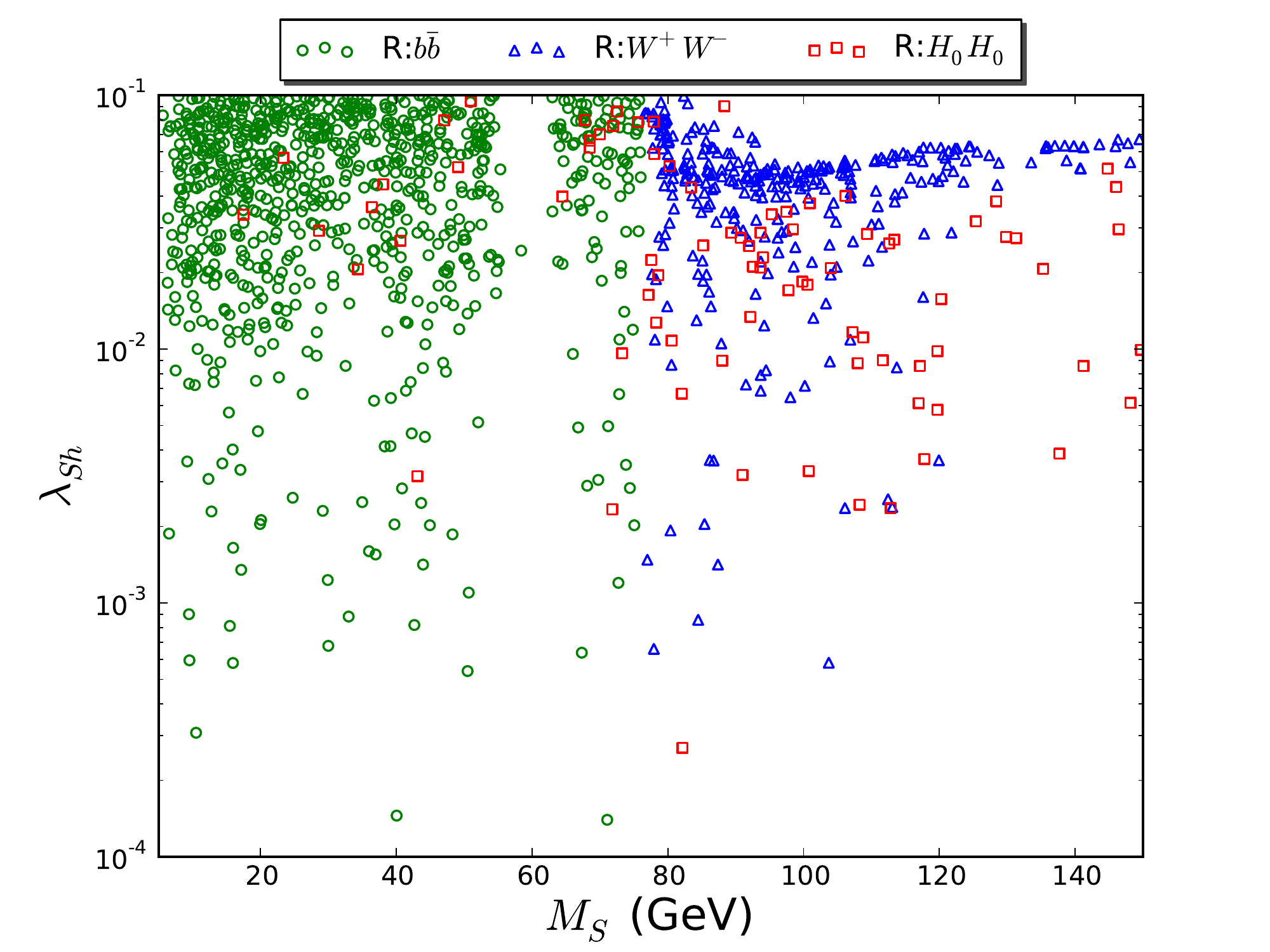}
\includegraphics[width=0.48\linewidth]{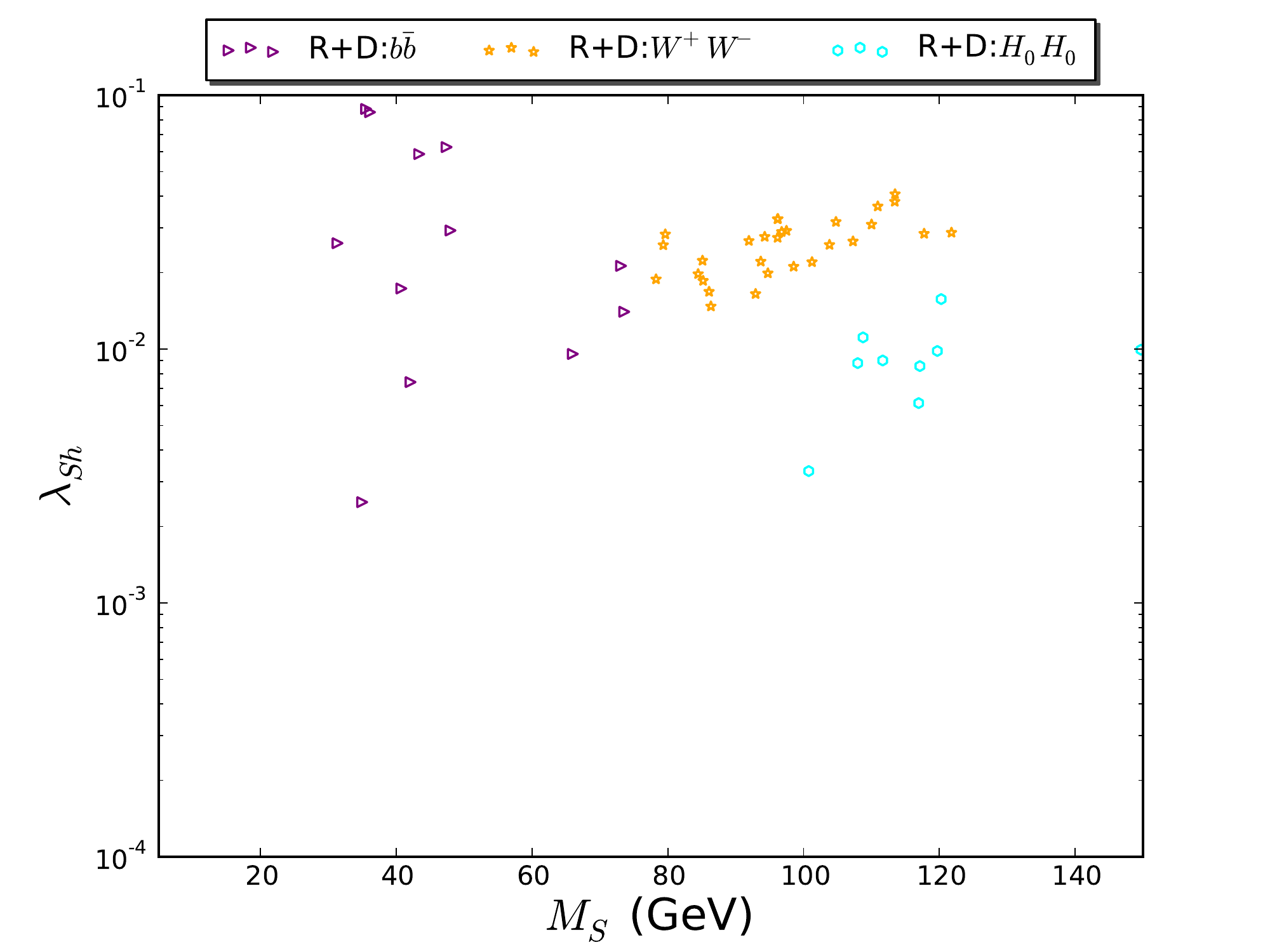}
\end{center}
\caption{Distribution of samples in dominant annihilation channels that survive R (left panel) or R+D (right) constraints is shown in the $M_S-\lambda_{Sh}$ plane for the low mass DM scenario.}
\label{fig:low1}
\end{figure}

\begin{figure}
\begin{center}
\includegraphics[width=0.48\linewidth]{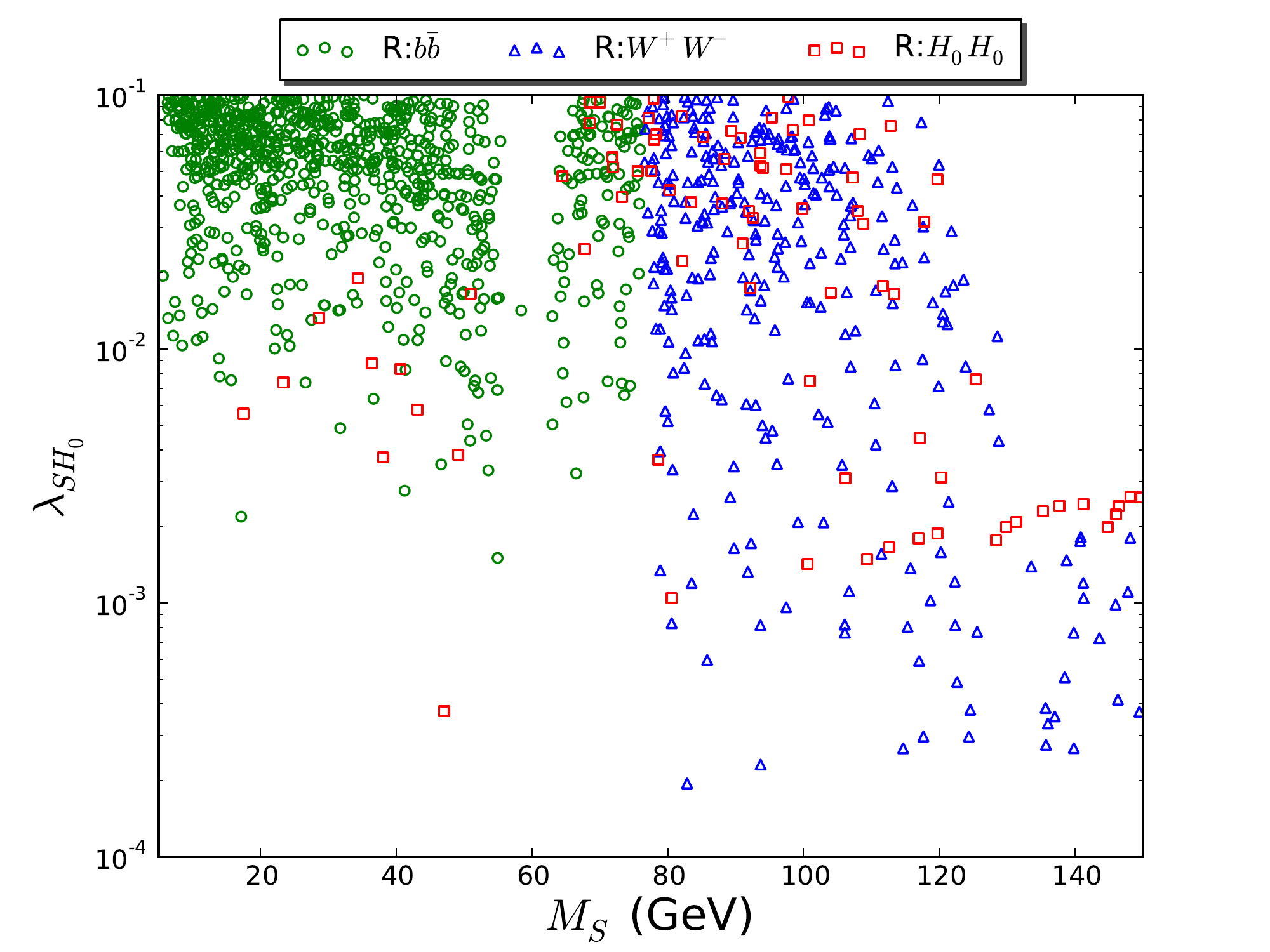}
\includegraphics[width=0.48\linewidth]{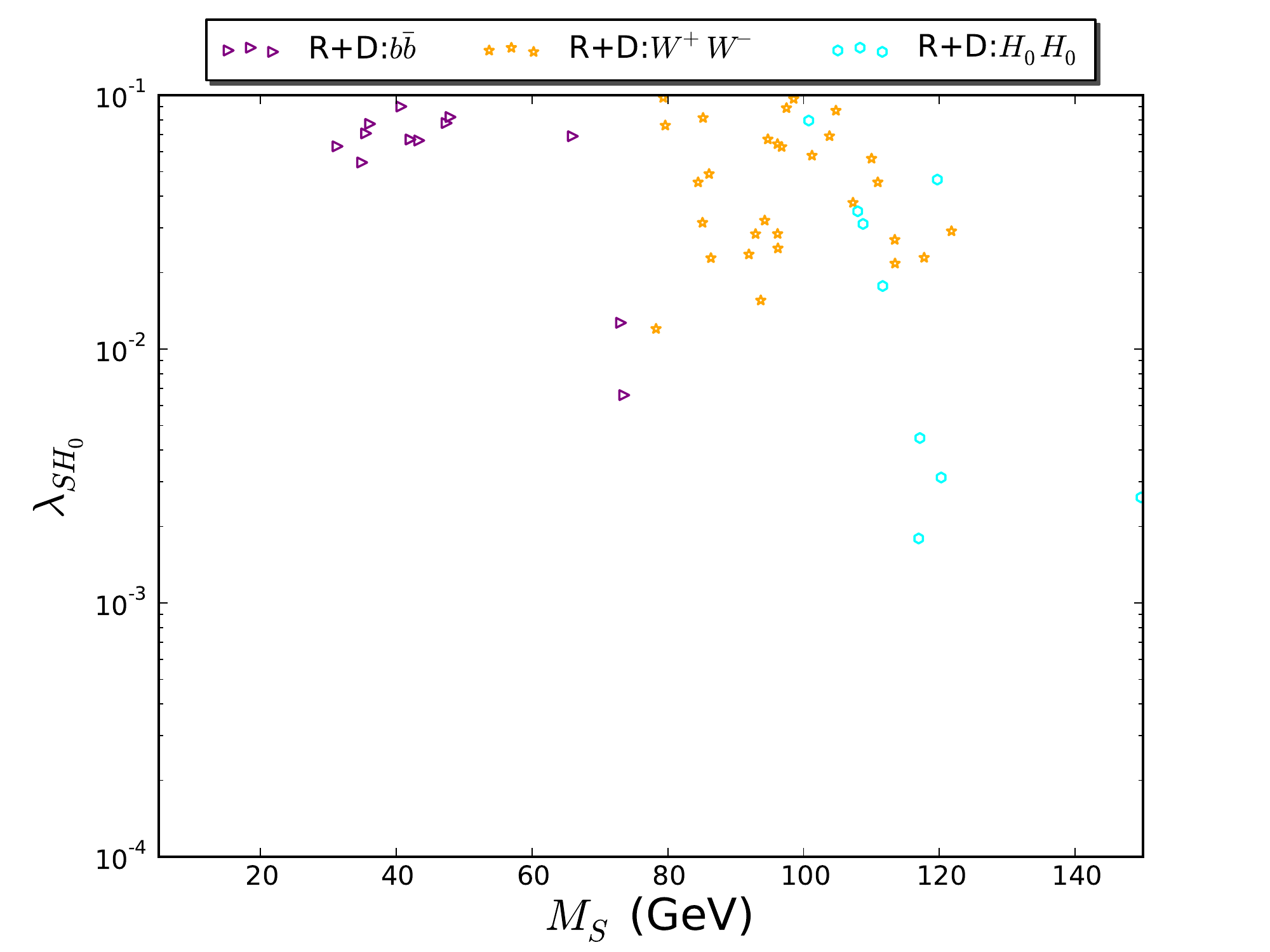}
\end{center}
\caption{Similar to Fig.~\ref{fig:low1} but in the $M_S-\lambda_{SH_0}$ plane.}
\label{fig:low2}
\end{figure}

\begin{figure}[!htbp]
\begin{center}
\includegraphics[width=0.48\linewidth]{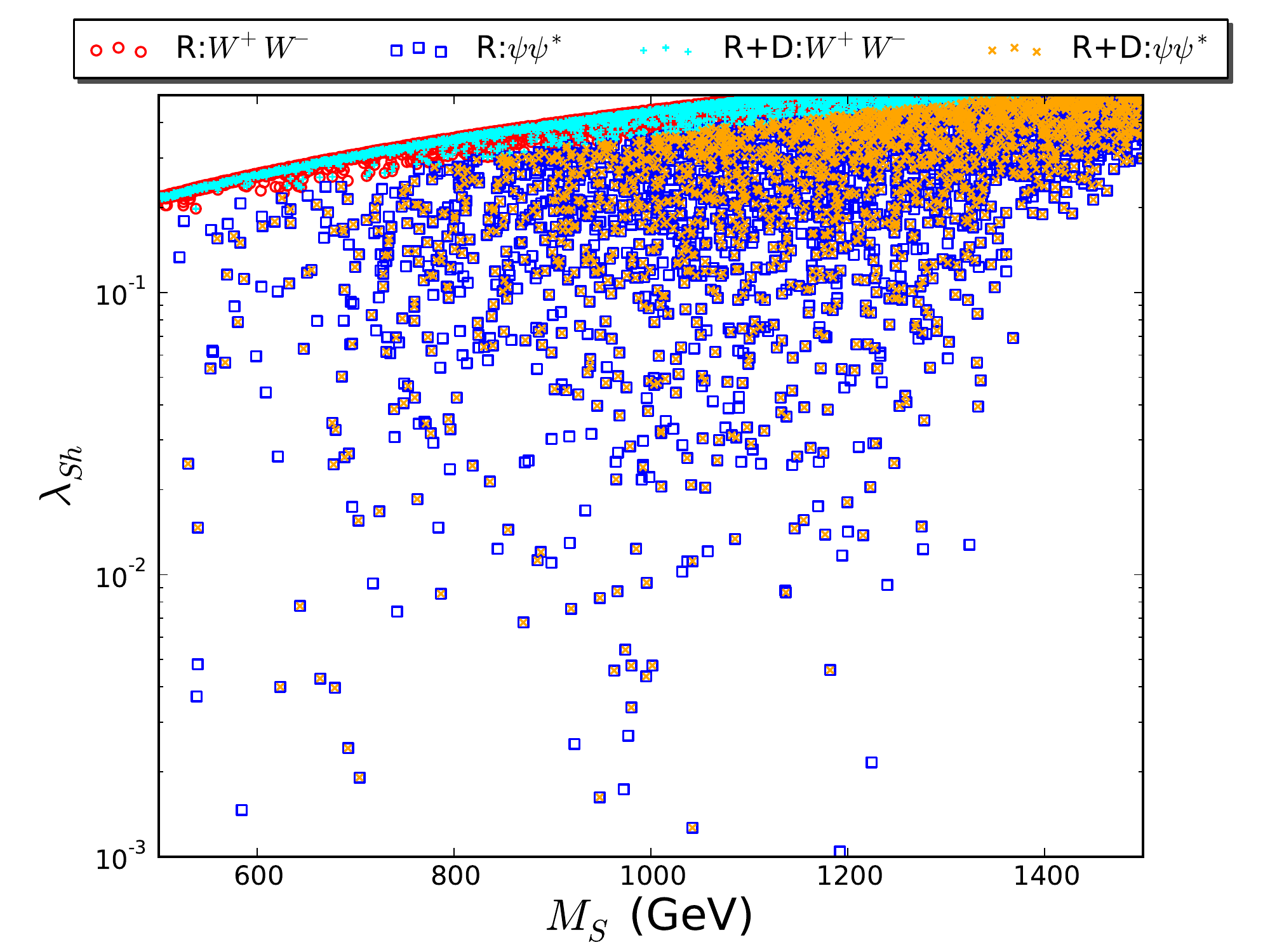}
\includegraphics[width=0.48\linewidth]{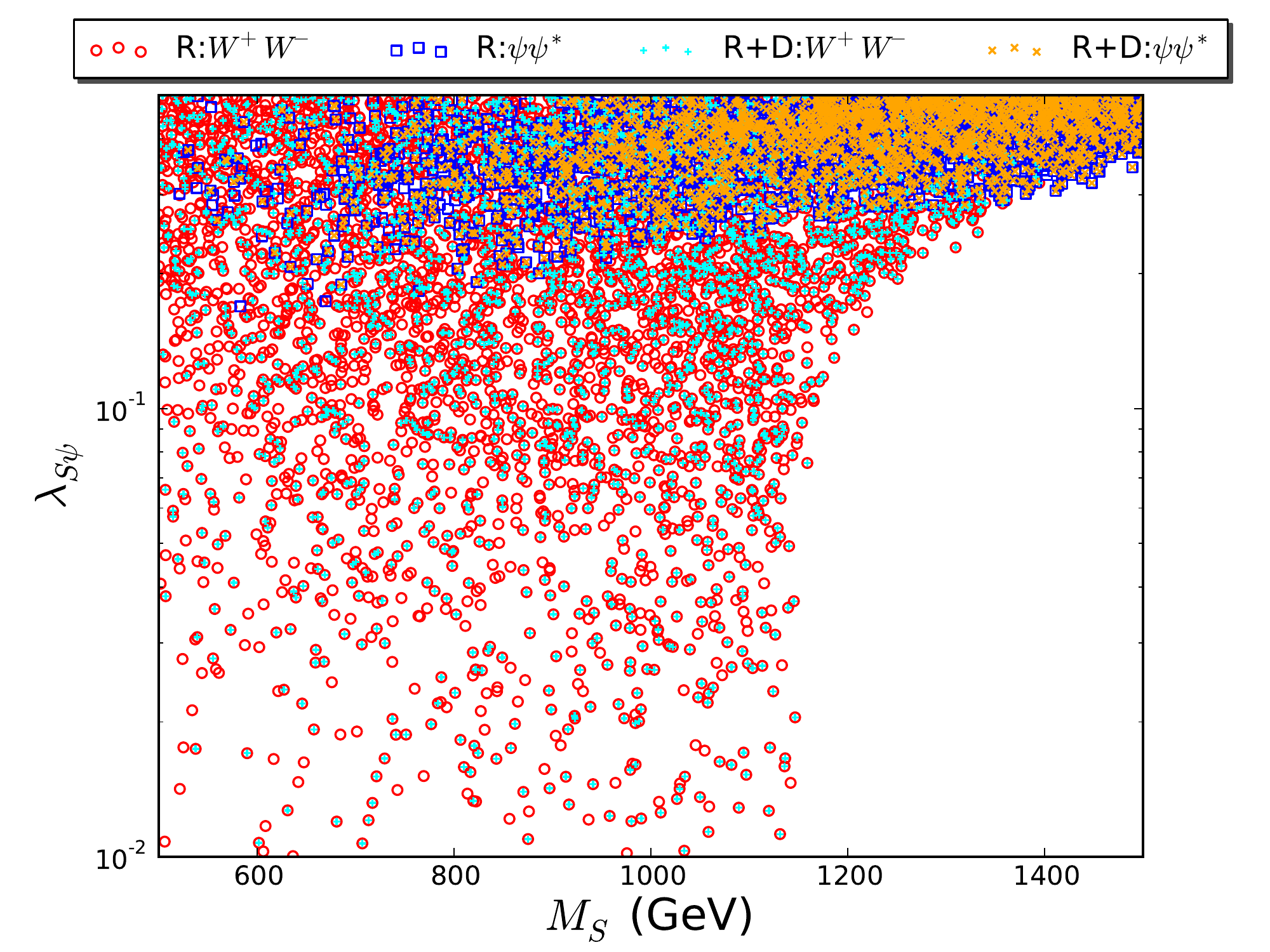}
\end{center}
\caption{Distribution of samples in dominant annihilation channels that survive R or R+D constraints is shown in the $M_S-\lambda_{Sh}$ (left) and $M_S-\lambda_\psi$ (right) plane for the high mass DM scenario.}
\label{fig:high1}
\end{figure}

For the purpose of illustrating the effects of various annihilation processes on relic abundance and direct detection, we list all important annihilation channels in Fig.~\ref{fig:channels}. A DM pair can annihilate into (1) a $b$ quark pair through the exchange of an $s$-channel $Z^\prime,~h,~H_0$, (2) an $H_0$ pair through their quartic interaction or via the exchange of an $s$-channel $h,~H_0$ or of a $t$-channel DM, (3) a $W$ boson pair via the exchange of an $s$-channel $h,~H_0$, and (4) a LQ pair via quartic interaction. We extract the dominant annihilation channel for each sample that survives relic abundance (R) alone or both relic abundance and direct detection (R+D). The distributions of survived samples are displayed for different projections of parameter space in Figs.~\ref{fig:low1} and \ref{fig:low2} in the low mass DM scenario and in Fig.~\ref{fig:high1} in the high mass DM scenario. For clarity, the number of survived samples in each dominant annihilation channel is listed in Table~\ref{tab:sample}. Several features learned from these results are summarized as follows.~\footnote{Notice that we have taken $M_{Z^\prime}=4~\TeV$ so that the relevant annihilation channel can be ignored for both scenarios.}

For the low mass DM scenario:
\begin{itemize}
\item There are much less survived samples than for the high mass scenario. This is due to the fact that the coupling between the DM and SM Higgs, $\lambda_{Sh}$, is tightly constrained by relic abundance and direct detections. As a consequence, only the channels mediated by the singlet scalar $H_0$ and $t$-channel DM can survive. On the contrary, the annihilation channels $SS^\ast\to W^+W^-/H_0 H_0$ are available in a wide parameter region. To be specific, regions of $\lambda_{Sh}\lesssim0.03$ and $\lambda_{SH_0}\gtrsim0.01$ are favoured for the low mass DM scenario.
\item The $b\bar{b}$ and $W^+W^-$ channels are respectively dominant when $M_S\lesssim75$ GeV and $\gtrsim75$ GeV. This is the usual behavior of the Higgs ($h/H_0$) portal DM~\cite{Cline:2013gha}. In addition, although the $H_0H_0$ channel could satisfy the relic abundance requirement in broad DM mass regions, only samples with $M_S>100~\GeV$ could escape direct detection bounds.
\end{itemize}

For the high mass DM scenario:
\begin{itemize}
\item Both $W^+W^-$ and $\psi\psi^\ast$ channels could be dominant when $M_S\lesssim 1.3~\TeV$, while only the $\psi\psi^\ast$ channel dominates for $M_S\gtrsim 1.3~\TeV$. The reason is that we have chosen the corresponding couplings $\lambda_{Sh},\lambda_{S\psi}<0.5$ in our scan.
\item $\lambda_{Sh}~(\lambda_{S\psi})\gtrsim 0.2$ is required when the $W^+W^-(\psi\psi^\ast)$ channel dominates. Moreover, the $W^+W^-$ channel fills a narrow band in the $\lambda_{Sh}-M_S$ plane where $\lambda_{Sh}$ increases with the increase of $M_S$, while the $\psi\psi^\ast$ channel in the same plane is much scattered.
\end{itemize}

\begin{table*}
\begin{tabular}{|c|c|c|c|c||c|c|c|}
\hline
& \multicolumn{4}{|c||}{Low mass DM ($3\times10^5$)} & \multicolumn{3}{|c|}{High mass DM ($3\times10^5$)}\\
\hline
Channels & total & $b\bar{b}$ & $W^{+}W^{-}$ & $H_0H_0$ & total & $W^{+}W^{-}$ & $\psi\psi^*$ \\
\hline
Relic (R) & $1216$ & $835$ & $300$ & $81$ & $6585$ & $3912$ & $2673$ \\\hline
Relic+Direct (R+D) & $50$ & $12$ & $29$ & $9$ & $4623$ & $2439$ & $2184$ \\\hline
\end{tabular}
\caption{Numbers of samples surviving R or R+D constraints for various dominant annihilation channels in low mass and high mass DM scenarios.}
\label{tab:sample}
\end{table*}

We now turn to GCE spectrum fitting in our model. The hard photons due to DM annihilation arise mainly from subsequent decays of SM particles, since their direct production is typically loop-suppressed. The continuous gamma-ray spectrum results from light mesons produced through hadronization and decay of SM fermions. The gamma-ray flux due to DM annihilation in the Galaxy can be expressed as
\begin{equation}
\frac{d\bm{\Phi}^\gamma}{dE_\gamma}=\frac{1}{4\pi}\frac{\bar{J}}{M^2_S}
\sum_f\langle\sigma v\rangle^f_{\rm halo}\frac{dN^\gamma_f}{dE_\gamma},\quad
\end{equation}
where $f$ sums over all quark and lepton annihilation channels. $\langle\sigma v\rangle^f_{\rm halo}$ is the thermally averaged annihilation cross section in the Galactic halo, and $dN^\gamma_f/dE_\gamma$ the prompt photon spectrum per annihilation for a given final state $f$. The astrophysical factor $\bar{J}$ is expressed as
\begin{equation}
\bar{J}=\int_{\Delta\Omega}d\Omega(b,l)\int_{\rm l.o.s}\rho^2(r(s,\bm{\psi}))ds,
\end{equation}
where $r(s,\bm{\psi})=\sqrt{r^2_\odot+s^2-2r_\odot s\cos\bm{\psi}}$. Here $r_\odot=8.5~\text{kpc}$ is the Sun-Galactic Center distance, $s$ is the line of sight (l.o.s) distance, and $\bm{\psi}$ is the angle between the observation direction and the Galactic Center. In terms of the Galactic latitude and longitude coordinate $(b,~l)$, one has
$\cos\bm{\psi}=\cos b\cos l$.

For a DM interpretation of the GCE, the angular region of interest for the Fermi-LAT is, $\Delta\Omega$: $2^\circ\leq|b|\leq 20^\circ$ and $|l|\leq20^\circ$. In our calculation, we take the generalized Navarro-Frenk-White (gNFW) profile for the DM halo distribution~\cite{Navarro:1995iw}
\begin{equation}
\rho (r)=\rho_\odot\left(\frac{r}{r_\odot}\right)^{-\gamma}
\left[\frac{1+r_\odot/r_s}{1+r/r_s}\right]^{3-\gamma}\;,
\end{equation}
where the scale radius $r_s=20~\text{kpc}$. Based on the analyses of Refs.~\cite{Agrawal:2014oha,Slatyer:2015jla,Berlin:2014pya,Basak:2014sza}, the local DM density $\rho_\odot$ and index $\gamma$ are estimated to be $\rho_\odot=(0.4\pm0.2)~{\rm GeV}/{\rm cm}^3$ and $\gamma=1.2\pm0.1$. We thus choose their central values $(\rho_\odot,~\gamma)=(0.4~{\rm GeV}/{\rm cm}^3,~1.2)$ for the benchmark halo profile, which yields the value $\bar{J}_{\rm ben}$ for $\bar{J}$. The uncertainties of $(\rho_\odot,~\gamma)$ then translate into $\bar{J}\equiv\mathcal{J}\bar{J}_{\rm ben}$, where the factor
$\mathcal{J}\in[0.14, 4.4]$ parameterizes the allowed range for DM distribution. We will do the GCE scan for $\mathcal{J}$ in the above range and $\mathcal{J}=1$ for the benchmark profile.

\begin{figure}[!htbp]
\begin{center}
\includegraphics[width=0.6\linewidth]{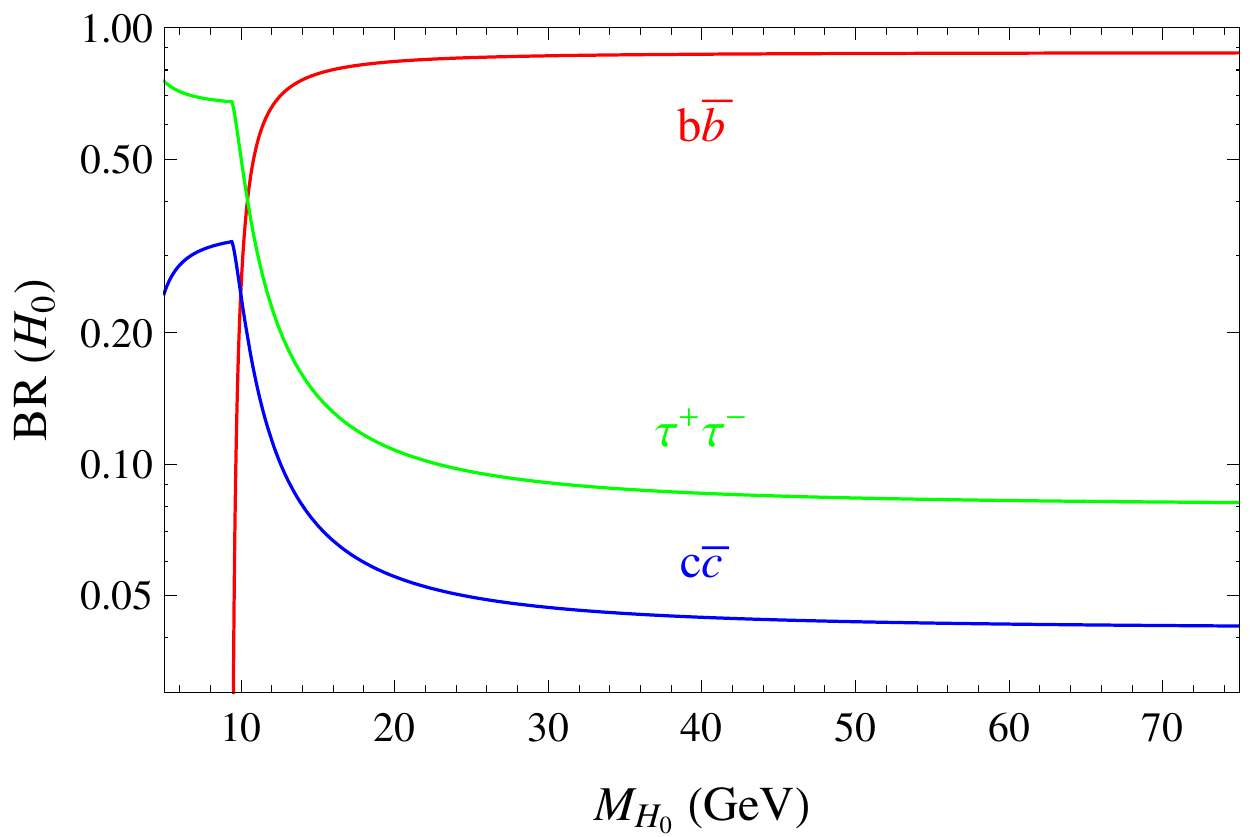}
\end{center}
\caption{Decay branching ratios of $H_0$ as a function of its mass.}
\label{fig:BR}
\end{figure}

To fit the GCE we use the results in Ref.~\cite{Calore:2014xka}, which explored in detail multiple galactic diffuse emission (GDE) models. We employ {\tt micrOMEGAs} and {\tt PPPC4DMID}~\cite{Cirelli:2010xx} to generate the photon spectrum and perform global fitting by using
\begin{equation}
\chi^2=\sum_{ij}\left(\frac{d\bm{\Phi}^{\rm th}_i}{dE_\gamma}
-\frac{d\bm{\Phi}^{\rm obs}_i}{dE_\gamma}\right)(\Sigma^{-1})_{ij}
\left(\frac{d\bm{\Phi}^{\rm th}_j}{dE_\gamma}
-\frac{d\bm{\Phi}^{\rm obs}_j}{dE_\gamma}\right)\;,
\end{equation}
where $d\bm{\Phi}^{\rm th,obs}_i/{dE_\gamma}$ are respectively the theoretical and observed gamma-ray flux in the $i$-th energy bin. $\Sigma_{ij}$ is the covariance matrix provided by Ref.~\cite{Calore:2014xka} which includes both statistical and correlated systematic errors. Here we focus on the on-shell mediator scenario, in which DM annihilates into a pair of on-shell singlet scalars $H_0$, which in turn decay to the SM quarks and leptons. The decay branching ratios of $H_0$ are presented in Fig.~\ref{fig:BR} versus its mass, which have a similar pattern to those of the SM Higgs due to the $\phi^0-\varphi^0$ mixing. We vary $M_S,~M_{H_0}$ in the GCE scan while fixing other parameters as shown in Table~\ref{tab:GCEscan}. In addition to relic abundance and direct detections, one must take into account the constraint from dwarf spheroidal galaxies (dSphs) in the Milky Way. The lack of gamma-ray excess from dSphs imposes a tight bound on the DM annihilation cross section in the galactic halo, and also gives a stringent constraint on the DM interpretation of GCE for various annihilation channels. Here we adopt dSphs limits provided in Ref.~\cite{Clark:2017fum}, which performed a model-independent and comprehensive analysis on various two-body and four-body annihilation channels based on the Planck~\cite{Ade:2015xua} (CMB), Fermi-LAT~\cite{GeringerSameth:2011iw,Ackermann:2011wa,Ackermann:2013yva,
Geringer-Sameth:2014qqa,Ackermann:2015zua} (dSphs) and AMS-02~\cite{Aguilar:2016vqr} (antiproton) results. For our model, most relevant are the $4b$, $4\tau$ and $2b2\tau$ channels. During the scan, we have translated corresponding limits into each $M_{H_0}$ sample weighted by $\br(H_0\to b\bar{b}/\tau^+\tau^-)$ and then extracted the most strict one.

\begin{figure}[!htbp]
\begin{center}
\includegraphics[width=0.44\linewidth]{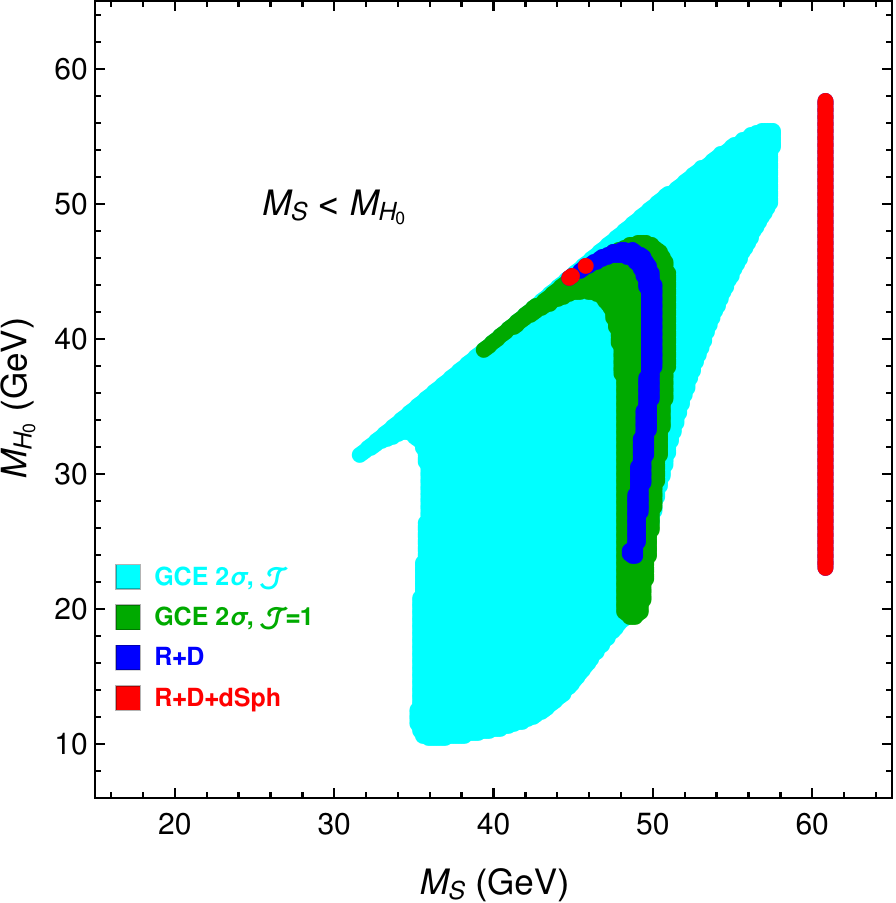}
\includegraphics[width=0.45\linewidth]{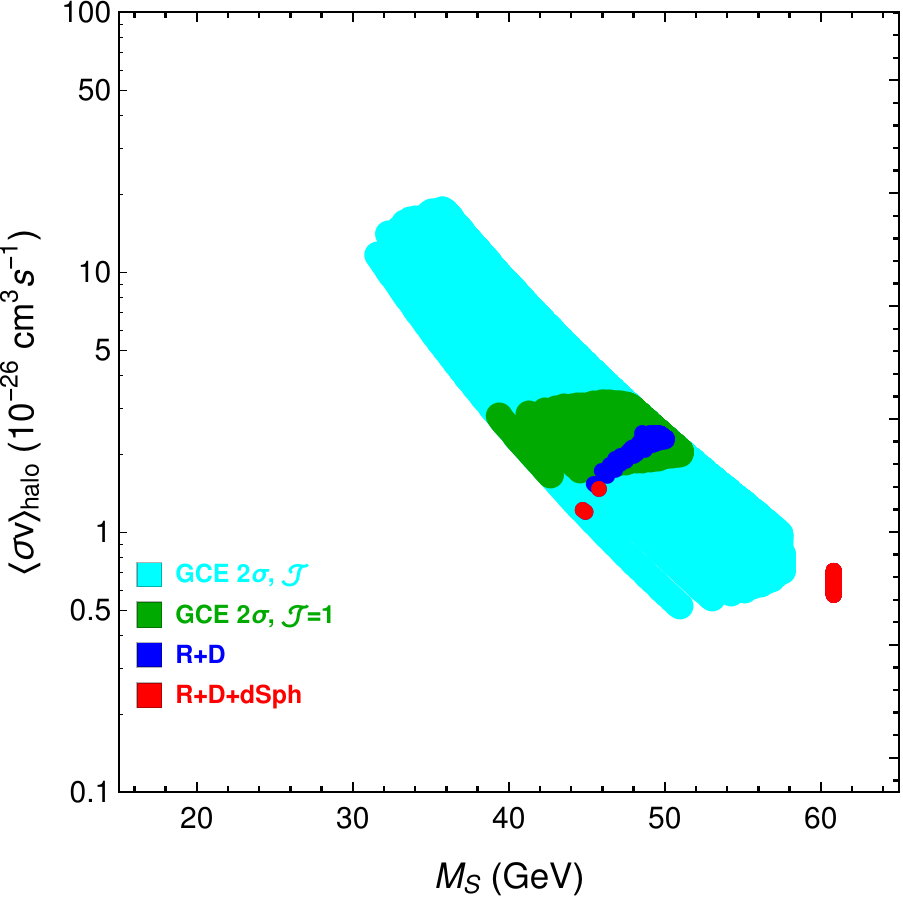}
\end{center}
\caption{Allowed regions for fitting GCE spectrum and various constraints in the $M_S-M_{H_0}$ (left) and $M_S-\langle\sigma v\rangle_{\rm halo}$ (right) plane. The cyan (green) region corresponds to GCE fitting for $\mathcal{J}\in [0.14,~4.4]$ ($\mathcal{J}=1$), while the blue (red) region satisfies R+D (R+D+dSph) constraints.}
\label{fig:GCE1}
\end{figure}

\begin{figure}[!htbp]
\begin{center}
\includegraphics[width=0.6\linewidth]{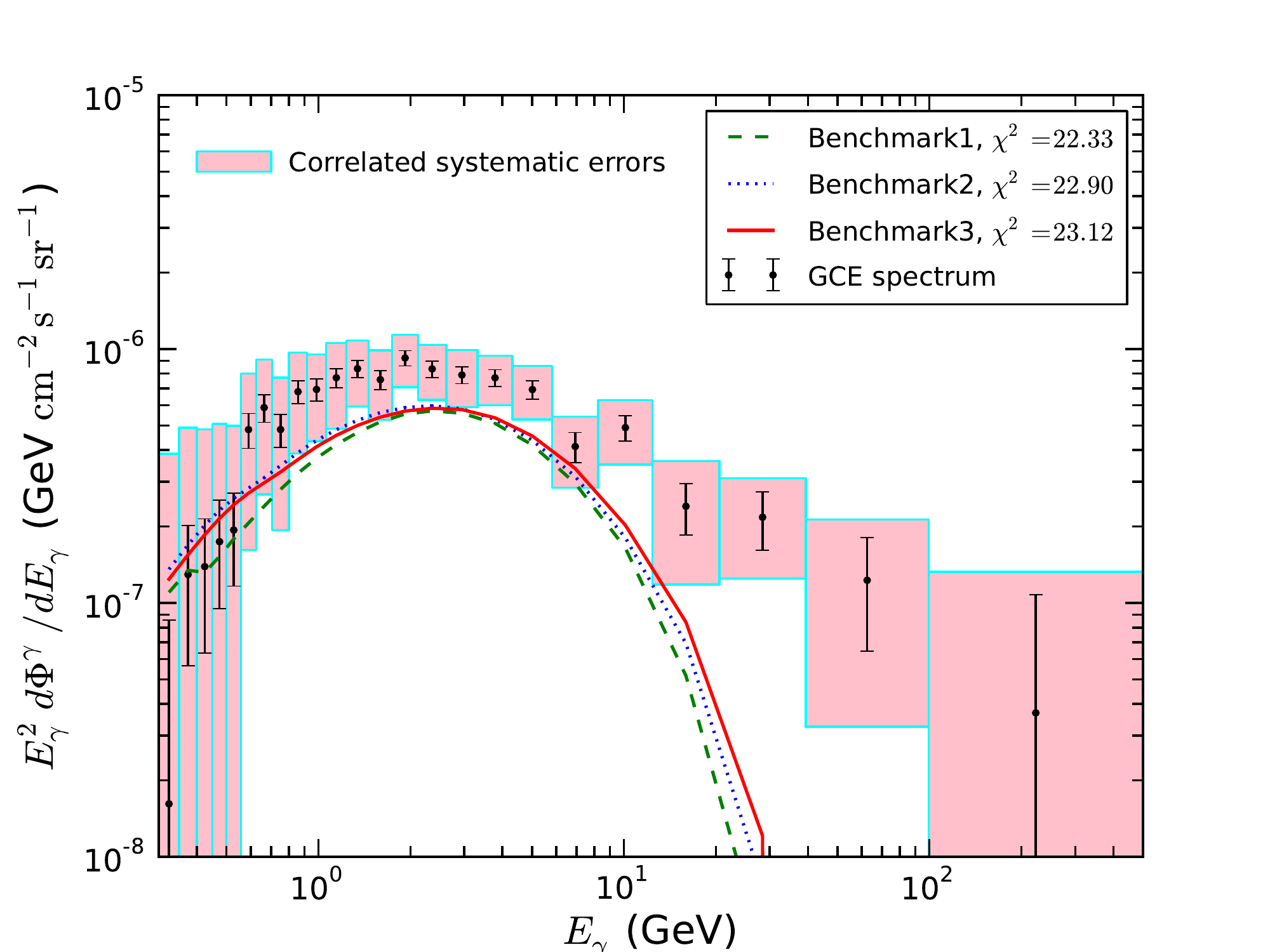}
\end{center}
\caption{The photon spectra for the three benchmarks in Table~\ref{tab:bench}. The GCE data with statistical and systematic errors (cyan ) in Ref.~\cite{Calore:2014xka}.}
\label{fig:flux}
\end{figure}

We present our results in Fig.~\ref{fig:GCE1}, where the allowed parameter regions for fitting the GCE spectrum and fulfilling various constraints are displayed in the $M_S-M_{H_0}$ (left panel) and $M_S-\langle\sigma v\rangle_{\rm halo}$ (right) plane. The cyan region corresponds to the $2\sigma$ ranges allowed by GCE fitting, i.e.,
for $\mathcal{J}\in [0.14,~4.4]$, and the green region is for the benchmark halo profile, i.e., $\mathcal{J}=1$. Scan samples that satisfy the R+D constraints cover the blue region, and those passing all of the R+D+dSph constraints are highlighted in red. Moreover, we show three benchmarks for GCE spectrum fitting in Fig.~\ref{fig:flux} and in Table~\ref{tab:bench}. Among them, the benchmark1 (benchmark2) is the best fit point of the GCE spectrum for $\mathcal{J}\in [0.14,~4.4]$ ($\mathcal{J}=1$) in the total samples, while benchmark3 is the best fit point in the R+D+dSph samples. Except for the benchmark1, the other two favor nearly degenerate $H_0$ and $S$ with $M_{H_0}\approx M_S\in[40,~50]~\GeV$. This feature can be understood
by a simple analysis of kinematics. For nearly degenerate $H_0$ and $S$, the $H_0$ pair is produced almost at rest and each decay final state of $H_0$ carries an energy $M_{H_0}/2\approx M_S/2$, which
results in a spectrum similar to the two-body annihilation process with a doubled number of injection fermions and reproduces the best fit result as the two-body $b\bar{b}$ final state. Finally, the exception of benchmark1 can be understood because it only occasionally gives a minimal $\chi^2$ by taking a marginal value of $\mathcal{J}$ and will yield an unacceptably large $\chi^2$ when fixing $\bar{J}=\bar{J}_{\rm ben}$.

\begin{table*}[hbtp]
\begin{tabular}{|c|c|c|c|c|c|c|c|c|c|c|c|}
\hline
&$M_S$ & $M_{H_0}$ & $M_{Z^\prime}$ & $g_{BL}$ & $\theta$ & $\lambda_{SH}$ & $\lambda_{SH_0}$ & $\lambda_{S \psi}$ & $M_\psi$ & $y^{ij}_L$ & $\mathcal{J}$\\\hline
GCE & $[5,75]$ & $[5,75]$ & $4000$ & $0.1$ & $10^{-2}$ & $10^{-3}$ & $10^{-3}$ & $0.5$ & $1000$ & $1$ & $[0.14, 4.4]$ or $1$\\\hline
\end{tabular}
\caption{The ranges or values of the input parameters used in GCE scan. All masses in units of GeV.}
\label{tab:GCEscan}
\end{table*}

\begin{table*}[hbtp]
\begin{tabular}{|c|c|c|c|c|c|c|c|c|}
\hline
& $M_S$ (GeV) & $M_{H_0}$ (GeV) & $\Omega_{\rm DM} h^2$ & $\sigma_{\rm SI}$ (${\rm cm}^2$) & $\langle\sigma v\rangle_{\rm halo}$ (${\rm cm}^3/{\rm s}$) & $\mathcal{J}$ & $\chi^2$ & R+D+dSph \\\hline
Benchmark1 & $36.61$ & $14.99$ & $0.023$  & $3.5\times10^{-45}$ & $1.28\times10^{-25}$ & $0.14$ & $22.33$ & Excluded \\\hline
Benchmark2 & $40.76$ & $40.59$ & $0.068$ & $1.05\times10^{-47}$ & $2.25\times10^{-26}$ & $1$ & $22.90$  &  Excluded  \\\hline
Benchmark3 & $44.74$ & $44.56$ & $0.1168$ & $2.97\times10^{-47}$ & $1.23\times10^{-26}$ & $2$ & $23.12$  & Allowed   \\\hline
\end{tabular}
\caption{Three benchmarks for GCE spectrum fitting. Here the benchmark1 (benchmark2) is the best fit point of GCE spectrum in the total samples for factor $\mathcal{J}\in [0.14,~4.4]$ ($\mathcal{J}=1$), and benchmark3 is the best fit point in the R+D+dSph samples.}
\label{tab:bench}
\end{table*}

\section{UHE neutrino events at IceCube}\label{sec:IceCube}

\subsection{Neutrino-nucleon scattering in SM and LQ contribution}
\label{sec:cs}

The IceCube neutrino observatory is located at the South Pole. The overwhelming majority events recorded by IceCube are muons from CR air showers, and only about one in a million events results from neutrino interactions. In the latter case, the UHE neutrinos in CR penetrate the ice and scatter with nucleons through neutrino-nucleon deep inelastic scattering (DIS) interaction. The Cherenkov light emitted by the secondary particles produced in scattering is observed by the IceCube detector. Depending on the interaction channel and incoming neutrino flavor, three types of signatures can be distinguished for neutrino events~\cite{Aartsen:2017sml}:
\begin{itemize}
  \item {The ``track-like" events, which are induced by muons produced in charged-current (CC) interactions of $\nu_\mu$.}
  \item {The ``shower-like" events, which are induced by neutral-current (NC) interactions of all neutrino flavors, and by CC interactions of $\nu_e$ in all energy ranges and $\nu_\tau$ with $E_{\nu_\tau}\leq 100~\TeV$.}
  \item {The ``double-bang" events, which are generated by high energy $\nu_\tau$. In this case its displaced vertices  between the hadronic shower at the $\tau$ generation and the shower produced at the $\tau$ decay can reach tens of meters.}
\end{itemize}
For the Yukawa structure in Eq.~(\ref{eq:yukawa2}) that we will employ for illustration, only ``track-like" CC and ``shower-like" NC events have to be taken into account in our calculation.

In the SM, the neutrino-nucleon ($\nu N$) interactions are mediated by the $W,~Z$ bosons:
\begin{eqnarray}
\nu_\ell+N &\to & \ell+X~~~~{\rm for~CC~interaction}, \label{cc1} \\
\nu_\ell+N & \to & \nu_\ell+X~~~{\rm for~NC~interaction}, \label{nc1}
\end{eqnarray}
where $\ell=e,\mu,\tau$ denotes the $SU(2)_L$ lepton flavor, $N=(n+p)/2$ is an isoscalar nucleon, and $X$ is the corresponding hadronic final state. At leading order (LO), the differential cross sections are~\cite{Gandhi:1995tf,Chen:2013dza}
\begin{eqnarray}
\frac{d^2\sigma_{\nu N}^{\rm CC}}{dxdy}&=& \frac{2G_F^2 M_NE_\nu}{\pi}
\frac{M_W^4}{(Q^2+M_W^2)^2}
\left[xf_q(x,Q^2)+xf_{\bar{q}}(x,Q^2)(1-y)^2\right],
\nonumber\\
\frac{d^2\sigma_{\nu N}^{\rm NC}}{dxdy}&=& \frac{G_F^2 M_N E_\nu}{2\pi}
\frac{M_Z^4}{(Q^2+M_Z^2)^2}
\left[xf_{q^0}(x,Q^2)+xf_{\bar{q}^0}(x,Q^2)(1-y)^2\right].
\label{eq:smnu}
\end{eqnarray}
In the above equations, $M_N$ and $M_{W,~Z}$ are respectively the nucleon and $W,~Z$ boson masses, $-Q^2$ is the momentum transfer squared, and $G_F$ is the Fermi constant. The Bjorken variables $x$ and $y$ are defined as,
\begin{eqnarray}
x=\frac{Q^2}{2M_NE_\nu y}\;,\quad\quad y=\frac{E_\nu-E_\ell}{E_\nu}\;, \label{xy}
\end{eqnarray}
where $E_\nu$ ($E_\ell$) is the energy of the incoming neutrino (outgoing lepton). The quark and anti-quark parton distribution functions (PDFs) $f_q,f_{\bar{q}}$ ($f_{q^0},f_{\bar{q}^0}$) are summed over all flavors of valence and sea quarks which are involved in CC (NC) interactions~\cite{Gandhi:1995tf,Chen:2013dza}:
\begin{eqnarray}
f_q &=& \frac{f_u+f_d}{2}+f_s+f_b\;,\nonumber\\
f_{\bar{q}} &=& \frac{f_{\bar{u}}+f_{\bar{d}}}{2}+f_c+f_t\;,\nonumber\\
f_{q^0} &=& \frac{f_u+f_d}{2}(L_u^2+L_d^2)+\frac{f_{\bar{u}}
+f_{\bar{d}}}{2}(R_u^2+R_d^2)+(f_s+f_b)(L_d^2+R_d^2)
+(f_c+f_t)(L_u^2+R_u^2)\;,\nonumber\\
f_{\bar{q}^0} &=& \frac{f_u+f_d}{2}(R_u^2+R_d^2)+\frac{f_{\bar{u}}
+f_{\bar{d}}}{2}(L_u^2+L_d^2)+(f_s+f_b)(L_d^2+R_d^2)
+(f_c+f_t)(L_u^2+R_u^2)\;,
\end{eqnarray}
where $R_d=(2/3)\sin^2\theta_W,~R_u=-2R_d,~L_d=-1+R_d$, and $L_u=1+R_u$ with $\theta_W$ the weak mixing angle. The cross sections for antineutrino-nucleon interactions ($\bar \nu N$) are obtained by the following replacements,
\begin{eqnarray}
\frac{d^2\sigma_{\bar{\nu} N}^{\rm CC,NC}}{dxdy} &=& \frac{d^2\sigma_{\nu N}^{\rm CC,NC}}{dxdy}(f_q\leftrightarrow f_{\bar{q}},~f_{q^0}\leftrightarrow f_{\bar{q}^0}).
\label{eq:smnubar}
\end{eqnarray}

The neutrino-electron interactions (in the target material) can generally be neglected compared to the neutrino-nucleon interactions due to the fact that $m_e\ll M_N$~\cite{Chen:2013dza}. The only important exception arises when the incoming neutrino has an energy of $E_\nu\sim 4-10$ PeV. In this case, the resonance production of the $W$ boson~\cite{Glashow:1960zz} enhances the $\bar{\nu}_e e$ cross section significantly with the peak at $E_\nu=M^2_W/2m_e=6.3$~PeV. Since this energy is higher than most of the shower events observed at IceCube, we do not include neutrino-electron interactions in our analysis; for a detailed discussion on this issue, see Ref.~\cite{Chen:2013dza}.

\begin{figure}[!htbp]
\begin{center}
\includegraphics[width=0.48\linewidth]{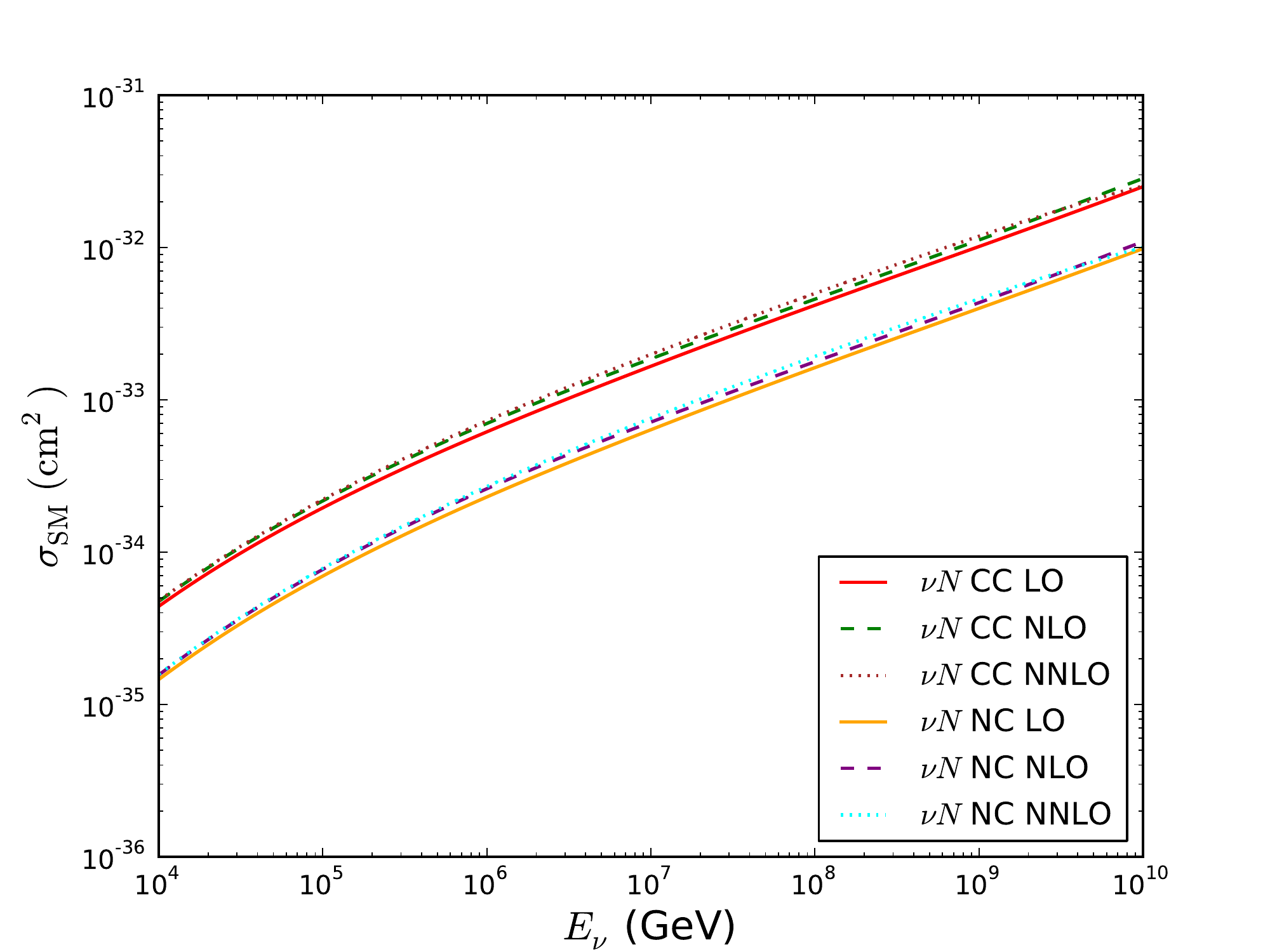}
\includegraphics[width=0.48\linewidth]{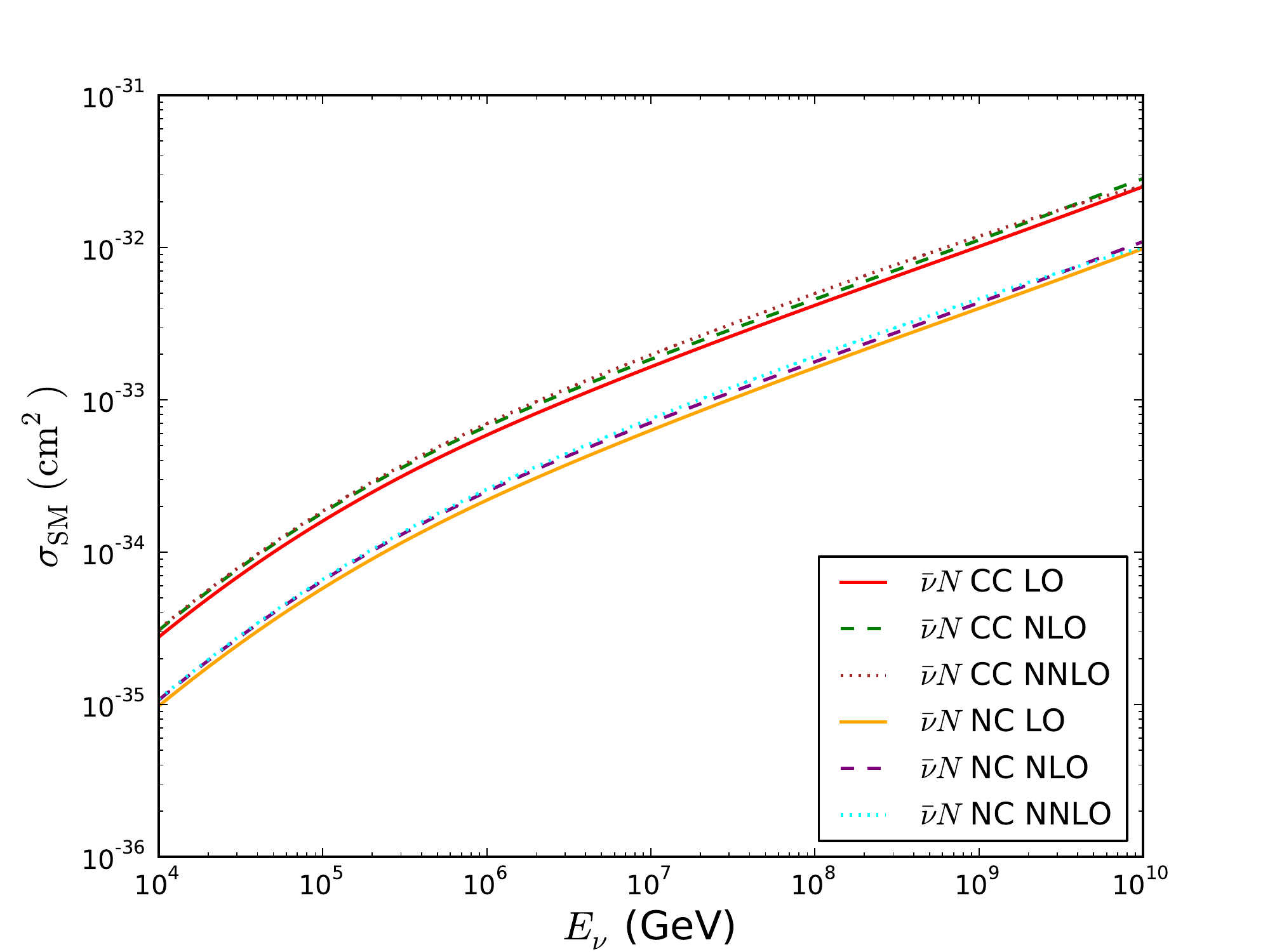}
\end{center}
\caption{Total $\nu N$ (left panel) and $\bar{\nu} N$ (right) scattering cross sections for the SM CC and NC processes as a function of neutrino energy $E_\nu$ with the PDFs at LO, NLO, and NNLO respectively.}
\label{fig:smtot}
\end{figure}

With differential cross sections in Eqs.~(\ref{eq:smnu}) and (\ref{eq:smnubar}), the total cross section is obtained by
\begin{equation}
\sigma(E_\nu)=\int^1_0\int^1_0dxdy\frac{d^2\sigma}{dxdy}\;.
\label{eq:smtot}
\end{equation}
In Fig.~\ref{fig:smtot}, we present the total SM cross section as a function of the incoming neutrino energy $E_\nu$ for both $\nu N$ and $\bar{\nu N}$ interactions using the {\tt NNPDF2.3} PDF sets~\cite{Ball:2012cx} at LO, NLO, and NNLO respectively. Due to the large uncertainty in small $x$ grids, we have set the lower limit of $x$ to be $10^{-6}$ in numerical integration to reach a reliable result, which is in good agreement with Ref.~\cite{Chen:2013dza}.

Now we compute the cross section due to LQ interactions. The neutrino-nucleon CC and NC processes are mediated by an $s$- and $u$-channel exchange of the LQ through Yukawa couplings in Eq.~(\ref{eq:yukawa}), and in addition there is interference between the LQ and SM amplitudes. Nevertheless we have numerically verified that both $u$-channel exchange and interference are negligible compared with the resonant $s$-channel LQ exchange. It is therefore sufficiently accurate to calculate the LQ contribution in the narrow width approximation (NWA) which only takes into account the $s$-channel resonance process. In order to keep at least two massive neutrinos as required by oscillation experiment, we assume a simple Yukawa structure in which only the first two generations of quarks and leptons are involved:
\begin{align}
y^{ij}_L=\left(
\begin{array}{ccc}
y_L^{11} & y_L^{12} & 0\\
y_L^{21} & y_L^{22} & 0\\
0 & 0 & 0
\end{array}\right)\quad {\rm and}\quad y^{ij}_R=0\;.
\label{eq:yukawa2}
\end{align}
In the NWA, the differential cross section for the NC or CC process can be written as
\begin{equation}
\frac{d\sigma^{\rm NC/CC}}{dy}(\nu_i N\to L_j X)=\frac{M_\psi}{32s\Gamma_\psi}\sum_{k,k^\prime}|y^{ik}_L|^2|y^{jk^\prime}_L|^2f_{q_k}(M^2_\psi/s,M^2_\psi y)\;,
\label{eq:csLQ1}
\end{equation}
where NC (CC) means $L_j=\nu_j~(\ell_j)$, $i,j,k,k^\prime=1,2$ refer to the first two generations of quarks and leptons, and $s=2M_NE_\nu$. Neglecting the final state fermion masses, the total decay width of the LQ $\psi$ is $\Gamma_\psi\simeq M_\psi/(8\pi)\sum_{ij}|y^{ij}_L|^2$. The Bjorken scaling variable $x$ has been integrated out in the NWA, so that the distribution functions are evaluated at $x=M^2_\psi/s$ and $Q^2=xys=M^2_\psi y$. The expressions for $\bar\nu N$ scattering can be obtained from Eq.~(\ref{eq:csLQ1}) by $f_{q}\leftrightarrow f_{\bar{q}}$.

\begin{figure}[!htbp]
\begin{center}
\includegraphics[width=0.6\linewidth]{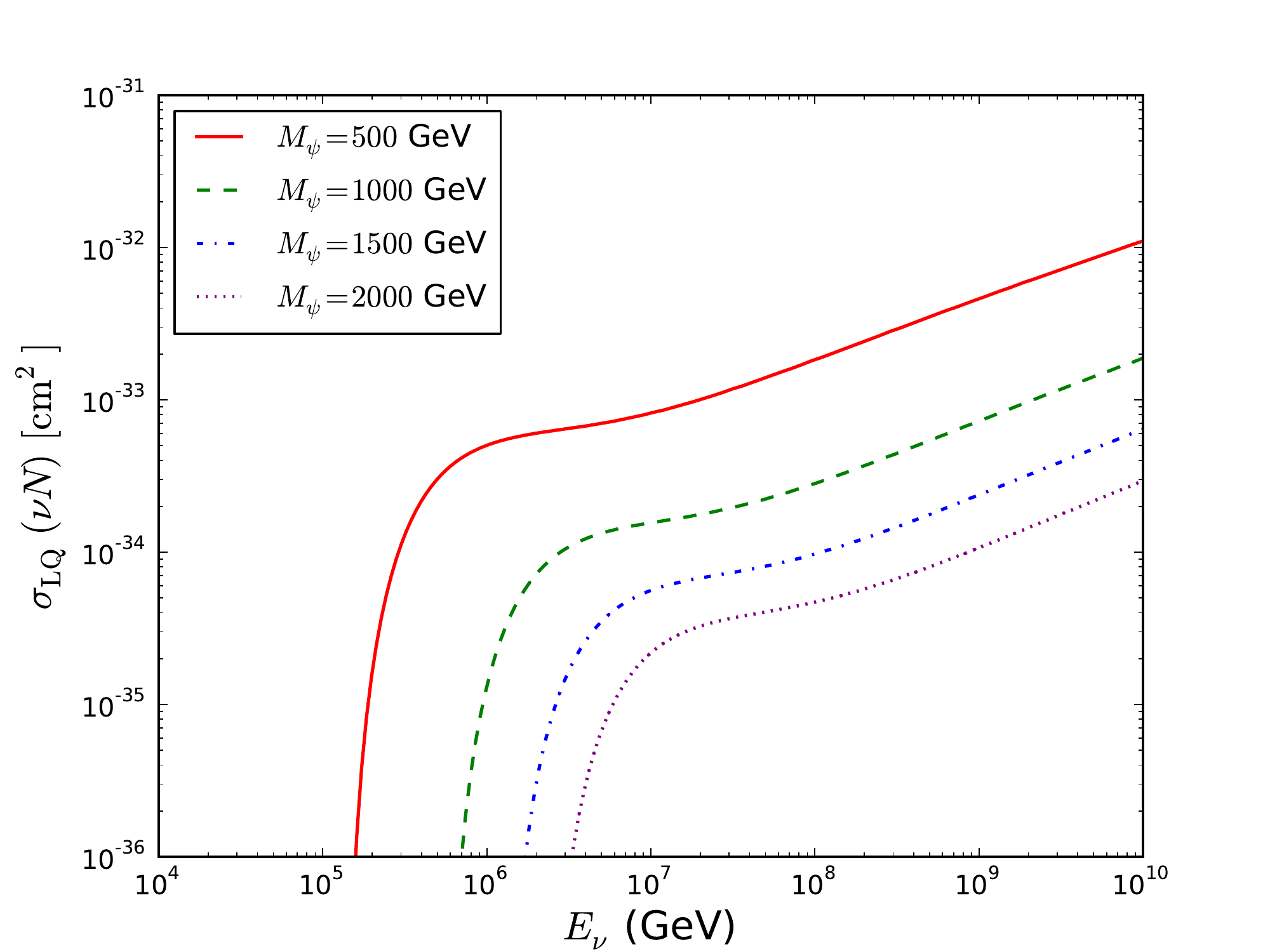}
\end{center}
\caption{Total $\nu N$ scattering cross section including the LQ CC and NC processes as a function of neutrino energy $E_\nu$ for typical values of $M_\psi$ and at $y^{11}_L=y^{21}_L=1$ and others vanishing.}
\label{fig:LQtot}
\end{figure}

For the purpose of illustration we plot in Fig.~\ref{fig:LQtot} the total $\nu N$ cross section due to the LQ resonance for typical values of $M_\psi$. We have assumed $y^{11}_L,~y^{21}_L=1$ and others vanishing, and included both NC and CC contributions. Comparing with the relatively smooth variation of the SM cross sections in Fig.~\ref{fig:smtot}, one finds that the LQ resonance contribution is triggered and rises rapidly once the incoming neutrino energy goes above the threshold $E^{\rm th}_\nu=M^2_\psi/(2M_N)$. Since $E^{\rm th}_\nu$ is in the multi TeV to PeV range in the current IceCube data, one expects that it is sensitive to the LQ in the mass range of $M_\psi\sim 100~\GeV-2~\TeV$. With the above preparation, we move on to evaluate the event rate at the IceCube which includes the LQ contribution and perform a statistical analysis to constrain model parameters.

\subsection{Event rate at IceCube and constraint on the model parameters}\label{sec:event}

The distribution of number of events with respect to the incoming neutrino energy and the inelasticity parameter is estimated as
\begin{equation}
\frac{dN}{dE_\nu dy}=T\cdot\Omega(E_\nu)\cdot N_{\rm eff}(E_\nu)\frac{d\bm{\Phi}^\nu}{dE_\nu}\frac{d\sigma}{dy}\;,
\label{eq:numnu}
\end{equation}
where $T$ is the exposure time, $\Omega(E_\nu)$ is the effective solid angle of coverage, $N_{\rm eff}(E_\nu)=N_A V_{\rm eff}(E_\nu)$ with $N_A=6.022\times10^{23}/{\rm cm}^3$ the water equivalent Avogadro number and $V_{\rm eff}(E_\nu)$ the effective target volume of the detector, $d\bm{\Phi}^\nu/dE_\nu$ the incoming neutrino flux, and $d\sigma/dy$ the differential $\nu N$ cross section shown in Eq.~(\ref{eq:csLQ1}) for the LQ contribution. In order to directly compare with IceCube data, one should use the electromagnetic (EM) equivalent deposited energy $E_{\rm dep}$ instead of the incoming neutrino energy $E_\nu$. For this purpose, we turn to calculate the expected number of events in a given EM equivalent deposited energy bin $[E^{\rm min}_{\rm dep},~E^{\rm max}_{\rm dep}]$ at IceCube, $N_{\rm bin}$, which can be expressed in terms of Eq.~(\ref{eq:numnu}) as follows,
\begin{eqnarray}
N_{\rm bin}
&=&\int^1_0dy\int^{E^{\rm max}_{\rm dep}}_{E^{\rm min}_{\rm dep}}dE_{\rm dep}\frac{dN}{dydE_{\nu}}\frac{dE_{\nu}}{dE_{\rm dep}}\nonumber\\
&=&T\int^1_0dy\int^{E_\nu(E^{\rm max}_{\rm dep},y)}_{E_\nu(E^{\rm min}_{\rm dep},y)}dE_\nu\; \Omega(E_\nu)\cdot N_{\rm eff}(E_\nu)\frac{d\bm{\Phi}^\nu}{dE_\nu}\frac{d\sigma}{dy}\;.
\label{eq:Nbin}
\end{eqnarray}

In the above equation, $E_{\rm dep}$ is always smaller than $E_\nu$ and their relation depends on the interaction channel. In this paper, we follow the method in Ref.~\cite{Chen:2013dza}. For NC events, the neutrino final state leads to missing energy, and the hadronic final state carries energy $E_X=y E_\nu$. Thus the total EM equivalent deposited energy for $\nu_{e,\mu}$ is given by
\begin{equation}
E^{\rm NC}_{\rm dep}=E_{\rm had}=F_X y E_\nu,
\label{eq:depNC}
\end{equation}
where the factor $F_X$ is the ratio of the number of photoelectrons originated from the hadronic shower to that from the equivalent-energy electromagnetic shower, which is a function of $E_X$ and parameterized as~\cite{Gabriel:1993ai}
\begin{eqnarray}
F_X=1-\left(\frac{E_X}{E_0}\right)^{-m}(1-f_0),
\end{eqnarray}
where the parameters $E_0,~m,~f_0$ are extracted from the simulations of a hadronic vertex cascade with the best-fit values $E_0=0.399$ GeV, $m=0.130$, and $f_0=0.467$ ~\cite{Kowalski:2004qc}. On the other hand, for CC events, the leptonic final states $e,~\mu$ entirely deposit their energy $E_{e,\mu}=(1-y)E_\nu$ into the EM shower. Together with the accompanying hadronic shower, the total EM equivalent deposited energy yields
\begin{equation}
E^{\rm CC}_{\rm dep}= E_{e,\mu}+E_{\rm had}.
\label{eq:depCC}
\end{equation}

The remaining parameters in Eq.~(\ref{eq:Nbin}) are determined as follows:
\begin{itemize}
\item Exposure time $T=2078~\text{days}$, corresponding to the IceCube data-taking period from year 2010 to 2016~\cite{Aartsen:2015zva}.
\item The effective target volume $V_{\rm eff}(E_\nu)=M_{\rm eff}/\rho_{\rm ice}$, where $\rho_{\rm ice}=0.9167~{\rm g}/{\rm cm}^3$ is the density of ice, and $M_{\rm eff}$ is the effective target mass. $M_{\rm eff}$ depends on the incoming neutrino energy and reaches the maximum value $\simeq 400$ Mton above 100 TeV for $\nu_e$ CC events (corresponding to $V_{\rm eff}^{\rm max}\simeq 0.44~{\rm km}^3$ water equivalent), and above 1 PeV for $\nu_{\mu,\tau}$ CC and NC events~\cite{Aartsen:2013jdh}. Here we choose $V_{\rm eff}=0.44~{\rm km}^3$ water equivalent in the calculation.
\item The solid angle of coverage $\Omega$ is different for neutrino events coming from the southern hemisphere (downgoing events) and northern hemisphere (upgoing events). While for isotropic downgoing events $\Omega$ is essentially equal to $2\pi$ sr, for isotropic upgoing events $\Omega$ is generally smaller by a shadow factor $S(E_\nu)$ due to the Earth attenuation effects~\cite{Gandhi:1995tf,Gandhi:1998ri}. The total solid angle of coverage is then given by $\Omega_{\rm tot}(E_\nu)=2\pi(1+S(E_\nu))~\text{sr}$. In the extreme case of a fully neutrino-opaque (neutrino-transparent) Earth, one has $\Omega_{\rm tot}=2\pi~\text{sr}$ ($4\pi~\text{sr}$), and for the realistic Earth one has $\Omega_{\rm tot}\in[2\pi,~4\pi]~\text{sr}$. The LQ could have a potential impact on the shadow factor through modification of the interaction length, but it has been shown in Ref.~\cite{Mileo:2016zeo} that this effect is small enough to be negligible. For simplicity, we will work with the above limiting values of $\Omega_{\rm tot}$ in our numerical analysis, and this will yield the two edges of the upper limit band on the Yukawa couplings $y^{ij}_L$ for a given LQ mass.
\item The incoming neutrino flux $d\bm{\Phi}^\nu/dE_\nu$ is assumed to be an isotropic, single power-law spectrum for each neutrino flavor $i$:
\begin{eqnarray}
\frac{d\bm{\Phi}^\nu_i}{dE_\nu} = \Phi_0f_i\left(\frac{E_\nu}{10^5 {\rm GeV}}\right)^{-\gamma}\;,
\label{eq:neuflux}
\end{eqnarray}
where $\Phi_0$ is the flux normalization at $E_\nu=10^5$ GeV for all neutrino flavors, $f_i$ is the fraction for the $i$th flavor at the Earth, and $\gamma$ the spectral index. Typical astrophysical processes yield source neutrinos with a flavor ratio of $\nu_e:\nu_\mu:\nu_\tau=1:2:0$ when they are produced by the decay of pions. Since the distance to the source is much larger than the neutrino oscillation length, one actually observes at the Earth an oscillation-averaged flavor composition, which tends to be in a ratio of $1:1:1$~\cite{Ahlers:2015lln}. We will thus use $f_i=1/3$ for $i=e,\mu,\tau$. For flux normalization $\Phi_0$ and spectral index $\gamma$, we assume the best-fit values in Ref.~\cite{Aartsen:2015knd}:
\begin{eqnarray}
\Phi_0=6.7^{+1.1}_{-1.2}\times 10^{-18}~{\rm GeV}
/({\rm sr~cm}^2~{\rm s})\;,~\gamma=2.50\pm0.09\;,
\label{eq:fnorm}
\end{eqnarray}
which were obtained by performing maximum likelihood combination of different IceCube results.
\end{itemize}

\begin{figure}[!htbp]
\begin{center}
\includegraphics[width=0.48\linewidth]{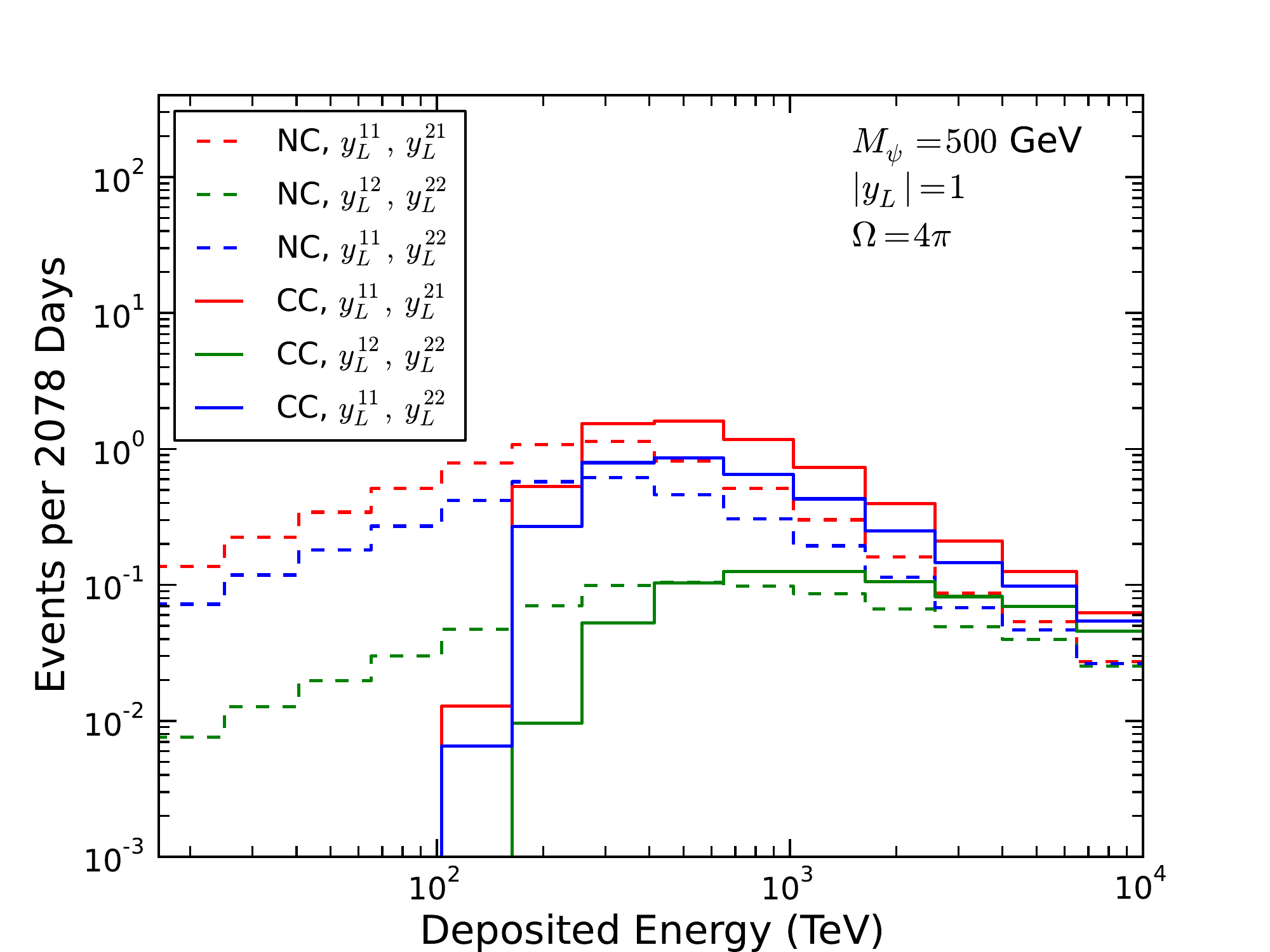}
\includegraphics[width=0.48\linewidth]{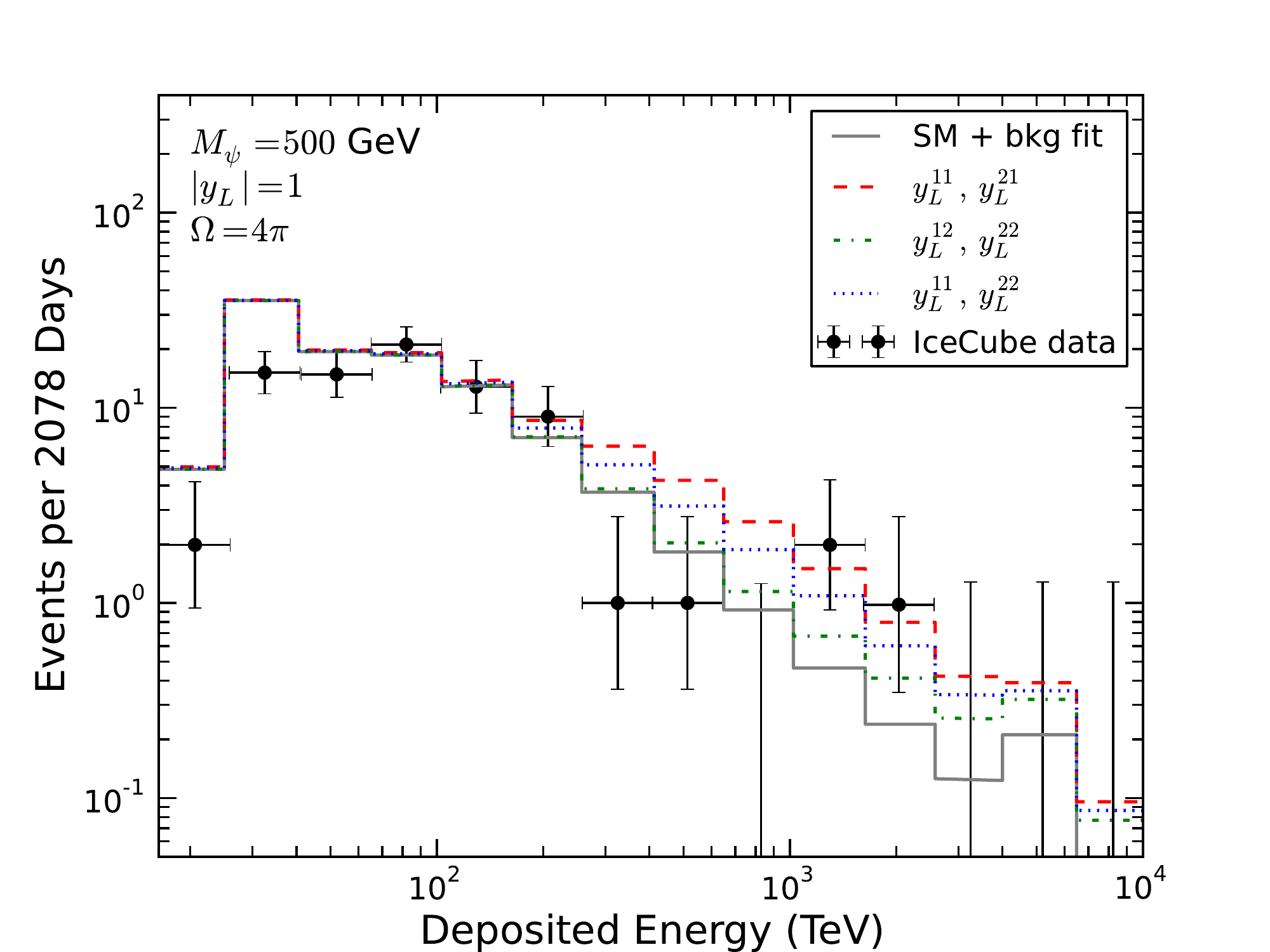}
\includegraphics[width=0.48\linewidth]{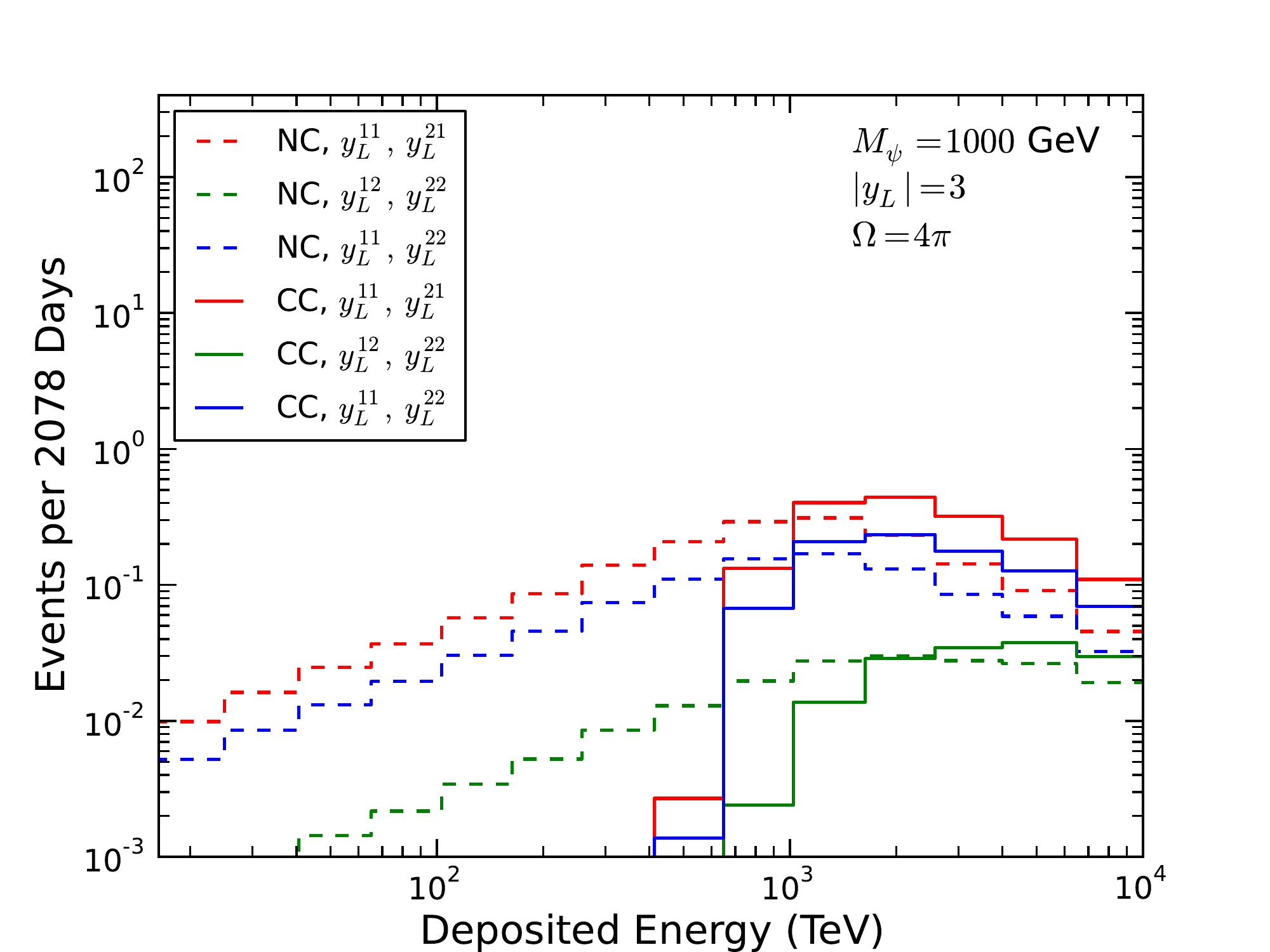}
\includegraphics[width=0.48\linewidth]{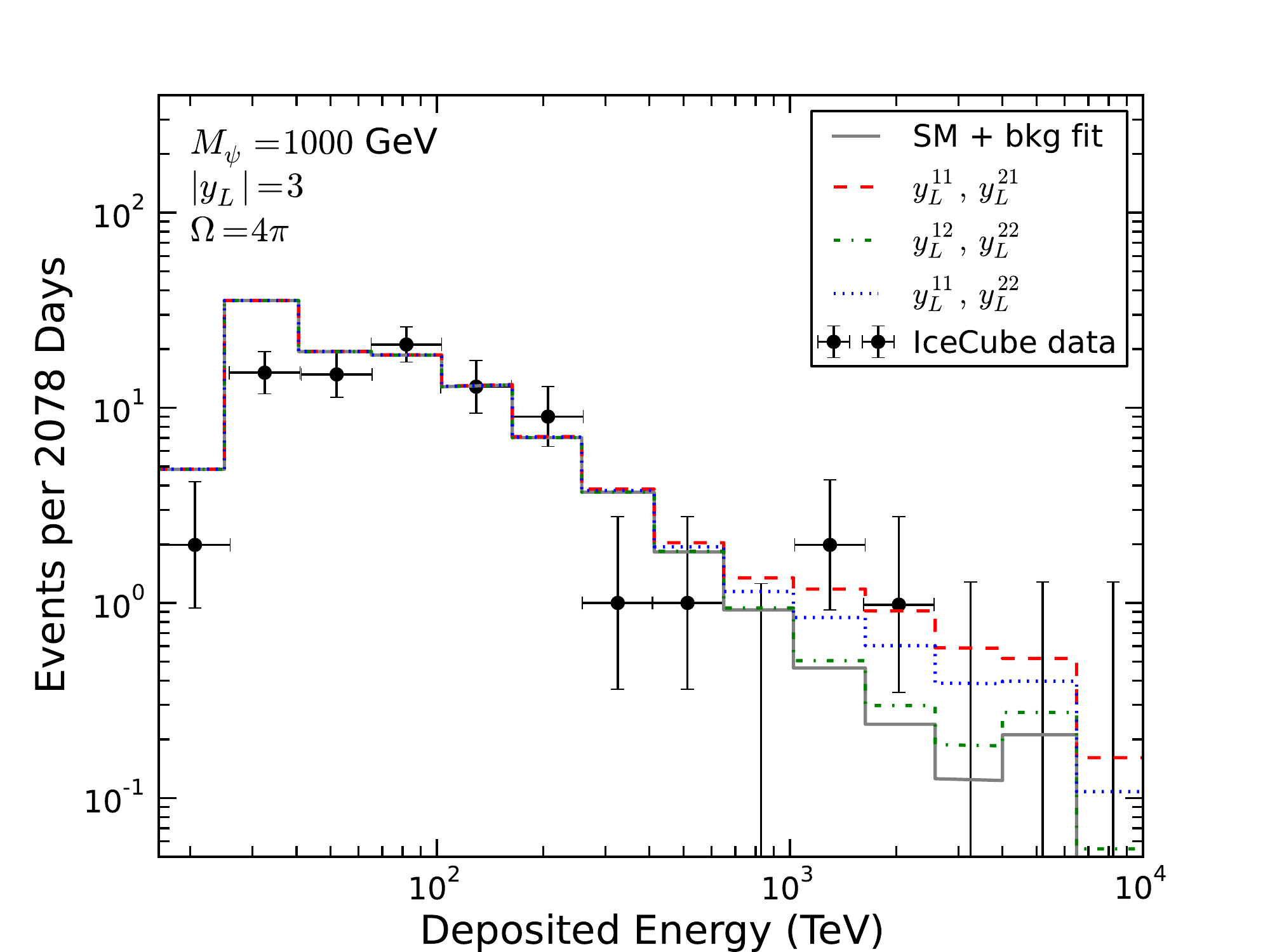}
\end{center}
\caption{Left panels: number of events due to pure LQ contribution as a function of deposited energy for various Yukawa components ($y^{11}_L,~y^{21}_L$), ($y^{12}_L,~y^{22}_L$), and ($y^{11}_L,~y^{22}_L$). Right panels: total numbers of events without LQ (SM + background, solid curve) and with LQ (SM + background + LQ) for the same Yukawa components (dashed, dash-dot and dotted curves) are compared with the 6-year IceCube data points. A universal value of $|y_L|=1~(3)$ is assumed for nonzero Yukawa components at $M_\psi=500~(1000)~\GeV$, and the solid angle of coverage is fixed at $\Omega=4\pi$.}
\label{fig:flux1}
\end{figure}

In order to investigate the number of events coming from the LQ contribution and its effect at the IceCube, we use Eq.~(\ref{eq:Nbin}) to calculate all of the 14 deposited energy bins in the IceCube data points. In the left panels of Fig.~\ref{fig:flux1}, we present the numbers of NC and CC events due to LQ as a function of the deposited energy. The plots are done for various Yukawa components in Eq.~(\ref{eq:yukawa2}) and typical LQ mass $M_\psi=500,~1000$ GeV, respectively. Here we simply assume a universal Yukawa coupling $|y_L|$ for the nonzero components and the legends in the figure are understood as follows: for instance, $(y^{11}_L,~y^{21}_L)$ indicates $y^{11}_L=y^{21}_L=|y_L|$ while others vanishing. It is straightforward to extend our analysis to non-universal cases by assuming specific relations for the Yukawa components in Eq.~(\ref{eq:yukawa2}). For comparison, the corresponding total numbers of events (SM + background + LQ) for the same Yukawa components and 6-year IceCube data points are presented in the right panels, where both IceCube data and SM + background fit are taken from Ref.~\cite{Aartsen:2017mau}. Some important information can be observed from Fig.~\ref{fig:flux1}:
\begin{itemize}
  \item The resonance peak broadens and shifts according to the threshold incoming neutrino energy $E^{\rm th}_\nu=M^2_\psi/(2M_N)$ for both NC and CC events.
  \item The CC events are distributed only in the deposited energy bins above the threshold energy, while the NC events are spread in all of bins. This arises from the fact that NC and CC processes deposit different amounts of energy according to Eqs.~(\ref{eq:depNC}, \ref{eq:depCC}), respectively.
  \item The numbers of events obey the sequence $N_{\rm bin}(y_L^{11},~y_L^{21})>N_{\rm bin}(y_L^{11},~y_L^{22})>N_{\rm bin}(y_L^{12},~y_L^{22})$, which clearly reflects the effects of PDF dependence. Since the $u$ and $d$ quarks are the dominant constituents of the nucleon, Yukawa components involving only the first generation of quarks give the most significant contribution while that involving the second generation of quarks is suppressed.
\end{itemize}

\begin{figure}[!htbp]
\begin{center}
\includegraphics[width=0.8\linewidth]{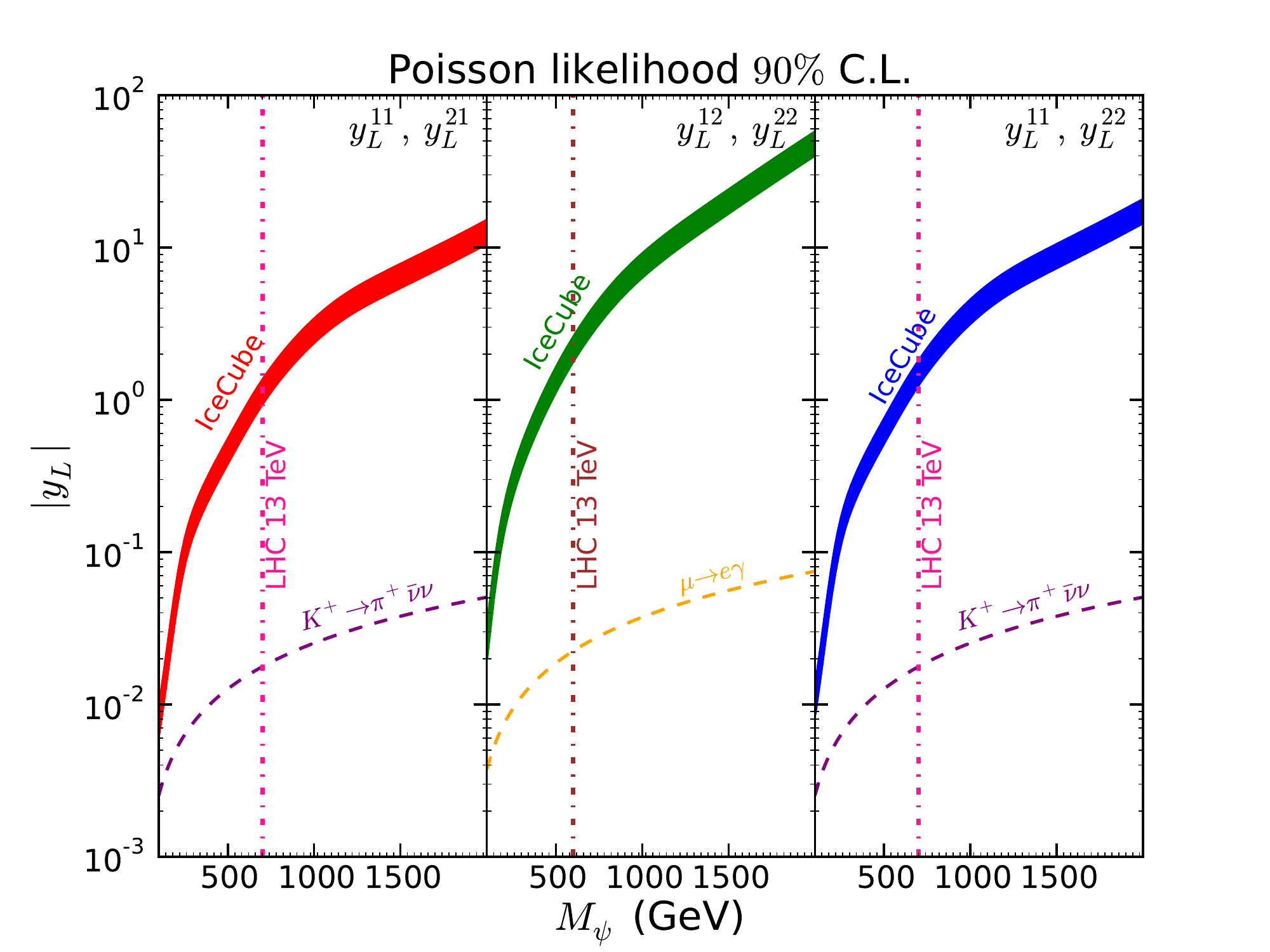}
\end{center}
\caption{$90\%$ C.L. upper limit bands (corresponding to a solid angle of coverage $\Omega\in[2\pi,~4\pi]$) on $|y_L|$ for various Yukawa structures versus LQ mass $M_\psi$ from the 6-year IceCube data. Also shown are $90\%$ C.L. limits from the decays $K^+\to\pi^+\bar{\nu}\nu$ (purple dashed lines) and $\mu\to e\gamma$ (orange dashed lines), and from direct searches at the 13 TeV LHC (magenta and brown vertical dot-dashed lines).}
\label{fig:excl}
\end{figure}

The interpretation of the IceCube excess in the energy interval $1-3$ PeV generically demands a LQ mass above TeV, where the production cross section and the neutrino flux are significantly suppressed. This may require a large Yukawa coupling beyond perturbation theory, for instance, $|y_L|=3$ for $M_\psi=1~\TeV$ as shown in the lower panels of Fig.~\ref{fig:flux1}. Nevertheless, one expects that a small fraction of the LQ contribution with a perturbative Yukawa coupling could relax the tension between the IceCube data and the SM prediction thus marginally improving the SM + background fit, which is also a part of motivation for this paper. Alternatively, one can also treat the current IceCube result as a complementary constraint, which allows us to put an upper bound on the Yukawa coupling for a given LQ mass. Along this way, we preform a binned statistical analysis with the Poisson likelihood function~\cite{Aartsen:2014muf,Dev:2016uxj},
\begin{equation}
L=\prod_{\rm bins}\frac{e^{-n^{\rm th}_i}\left(n^{\rm th}_i\right)^{n^{\rm obs}_i}}{n^{\rm obs}_i!}\;,
\end{equation}
where $n^{\rm obs,~th}_i$ are respectively the observed and theory counts in the $i$-th bin. We then use the test statistics
\begin{equation}
-2\Delta \ln L=-2(\ln L -\ln L_{\rm max})\;,
\end{equation}
to derive upper limits on $y_L^{ij}$ at $90\%$ C.L. (corresponding to $-2\Delta \ln L=2.71$) in the LQ mass region $M_\psi\in[100,~2000]~\GeV$. Here $L_{\rm max}$ is the likelihood value assuming $y_L^{ij}=0$. Our results are presented in Fig.~\ref{fig:excl} for the same Yukawa structure discussed above. As expected, the most stringent bound is set on the $(y^{11}_L,~y^{21}_L)$ components, while that on $(y^{12}_L,~y^{22}_L)$ is relatively weak due to subdominant PDFs of the second generation of quarks in the proton.

There also exist stringent limits on $y_L^{ij}$ from flavor physics and on $M_\psi$ from LHC direct searches. For the former, according to our discussion in section~\ref{sec:flavor}, the components $(y^{11}_L,~y^{21}_L)$ and $(y^{11}_L,~y^{22}_L)$ components are most sensitive to the $K$-meson decay $K^+\to\pi^+\bar{\nu}\nu$, while $(y^{12}_L,~y^{22}_L)$ are sensitive to the LFV decay $\mu\to e\gamma$. As an illustration of the collider constraints, we use the ATLAS limits on the LQ mass at 13 TeV~\cite{Aaboud:2016qeg}. These limits are also shown in Fig.~\ref{fig:excl} for comparison. In all the cases, the limits derived from $K^+\to\pi^+\bar{\nu}\nu$ and $\mu \to e \gamma$ decays are much stronger than that from the IceCube in the entire mass range considered. This severely restricts the LQ interpretation of the IceCube excess in the 6-year data. However, it is worthwhile to treat the excess as a supplementary constraint although it is highly limited by current statistics. With the increase of exposure time and data collection, one expects that the IceCube limit will improve and that the distribution of data in the bins may even change remarkably. In that case better agreement or more severe discrepancy with the SM prediction will serve as a complementary limit or hint of new physics.

\section{Conclusion}\label{sec:con}

We have investigated the phenomenology of the colored Zee-Babu model augmented with a $U(1)_{B-L}$ gauge symmetry and a singlet scalar DM $S$. The tiny neutrino masses are still generated via a two-loop radiative seesaw involving the SM quarks, a diquark and a LQ, but now we have made connections to two high energy CR observations: the Fermi-LAT GCE and the PeV UHE neutrino events at the IceCube. For the Fermi-LAT GCE, we focused on the annihilation channel in which the singlet (-dominating) Higgs $H_0$ acts as an on-shell mediator. We found that the GCE spectrum is well fitted when the $H_0$ mass is close to the DM mass which is consistent with constraints coming from relic abundance, direct detections as well as dSphs in the Milky Way. We studied the feasibility that the resonance LQ production is responsible for the extra UHE neutrino events at the IceCube. Using the 6-year dataset in the multi TeV to PeV energy range, we derived upper limits on the LQ Yukawa couplings as a function of its mass. Although the fraction of the LQ contribution to the IceCube excess is tightly limited by flavor physics constraints at low energies, we expect that better limits will be possible with more statistics in the near future. Together with the limits from LHC direct searches, the parameter space will be explored complementarily by multi-experiments.

\section*{Acknowledgements} \label{acknowledgements}

We thank Qing-Hong Cao, Yandong Liu, Donglian Xu, Yi-Lei Tang, Jue Zhang and Yang Zhang for help and useful discussions on various aspects of this paper. This work was supported in part by the Grants No. NSFC-11575089 and No. NSFC-11025525, by The National Key Research and Development Program of China under Grant No. 2017YFA0402200, and by the CAS Center for Excellence in Particle Physics (CCEPP).



\begin{thebibliography}{000}


\bibitem{Goodenough:2009gk}
  L.~Goodenough and D.~Hooper,
  arXiv:0910.2998 [hep-ph].

\bibitem{Hooper:2010mq}
  D.~Hooper and L.~Goodenough,
  Phys.\ Lett.\ B {\bf 697} (2011) 412
  [arXiv:1010.2752 [hep-ph]].

\bibitem{Hooper:2011ti}
  D.~Hooper and T.~Linden,
  Phys.\ Rev.\ D {\bf 84} (2011) 123005
  [arXiv:1110.0006 [astro-ph.HE]].

\bibitem{Abazajian:2012pn}
  K.~N.~Abazajian and M.~Kaplinghat,
  Phys.\ Rev.\ D {\bf 86} (2012) 083511
   Erratum: [Phys.\ Rev.\ D {\bf 87} (2013) 129902]
  [arXiv:1207.6047 [astro-ph.HE]].

\bibitem{Gordon:2013vta}
  C.~Gordon and O.~Macias,
  Phys.\ Rev.\ D {\bf 88} (2013) no.8,  083521
   Erratum: [Phys.\ Rev.\ D {\bf 89} (2014) no.4,  049901]
  [arXiv:1306.5725 [astro-ph.HE]].

\bibitem{Abazajian:2014fta}
  K.~N.~Abazajian, N.~Canac, S.~Horiuchi and M.~Kaplinghat,
  Phys.\ Rev.\ D {\bf 90} (2014) no.2,  023526
  [arXiv:1402.4090 [astro-ph.HE]].

\bibitem{Daylan:2014rsa}
  T.~Daylan, D.~P.~Finkbeiner, D.~Hooper, T.~Linden, S.~K.~N.~Portillo, N.~L.~Rodd and T.~R.~Slatyer,
  Phys.\ Dark Univ.\  {\bf 12} (2016) 1
  [arXiv:1402.6703 [astro-ph.HE]].

\bibitem{Calore:2014xka}
  F.~Calore, I.~Cholis and C.~Weniger,
  JCAP {\bf 1503} (2015) 038
  [arXiv:1409.0042 [astro-ph.CO]].

\bibitem{Yuan:2014rca}
  Q.~Yuan and B.~Zhang,
  JHEAp {\bf 3-4} (2014) 1
  [arXiv:1404.2318 [astro-ph.HE]].

\bibitem{Bartels:2015aea}
  R.~Bartels, S.~Krishnamurthy and C.~Weniger,
  Phys.\ Rev.\ Lett.\  {\bf 116} (2016) no.5,  051102
  [arXiv:1506.05104 [astro-ph.HE]].

\bibitem{Lee:2015fea}
  S.~K.~Lee, M.~Lisanti, B.~R.~Safdi, T.~R.~Slatyer and W.~Xue,
  Phys.\ Rev.\ Lett.\  {\bf 116} (2016) no.5,  051103
  [arXiv:1506.05124 [astro-ph.HE]].

\bibitem{TheFermi-LAT:2017vmf}
  M.~Ackermann {\it et al.} [Fermi-LAT Collaboration],
  Astrophys.\ J.\  {\bf 840}, no. 1, 43 (2017)
  [arXiv:1704.03910 [astro-ph.HE]].

\bibitem{Fermi-LAT:2017yoi}
  M.~Ajello {\it et al.} [Fermi-LAT Collaboration],
  [arXiv:1705.00009 [astro-ph.HE]].

\bibitem{Haggard:2017lyq}
  D.~Haggard, C.~Heinke, D.~Hooper and T.~Linden,
  JCAP {\bf 1705}, no. 05, 056 (2017)
  [arXiv:1701.02726 [astro-ph.HE]].

\bibitem{Bartels:2017xba}
  R.~Bartels, D.~Hooper, T.~Linden, S.~Mishra-Sharma, N.~L.~Rodd, B.~R.~Safdi and T.~R.~Slatyer,
  arXiv:1710.10266 [astro-ph.HE].

\bibitem{Agrawal:2014oha}
  P.~Agrawal, B.~Batell, P.~J.~Fox and R.~Harnik,
  JCAP {\bf 1505}, 011 (2015)
  [arXiv:1411.2592 [hep-ph]].

\bibitem{Cline:2015qha}
  J.~M.~Cline, G.~Dupuis, Z.~Liu and W.~Xue,
  Phys.\ Rev.\ D {\bf 91}, no. 11, 115010 (2015)
  [arXiv:1503.08213 [hep-ph]].

\bibitem{Elor:2015tva}
  G.~Elor, N.~L.~Rodd and T.~R.~Slatyer,
  Phys.\ Rev.\ D {\bf 91}, 103531 (2015)
  [arXiv:1503.01773 [hep-ph]].

\bibitem{Calore:2014nla}
  F.~Calore, I.~Cholis, C.~McCabe and C.~Weniger,
  Phys.\ Rev.\ D {\bf 91}, no. 6, 063003 (2015)
  [arXiv:1411.4647 [hep-ph]].

\bibitem{Cirelli:2014lwa}
  M.~Cirelli, D.~Gaggero, G.~Giesen, M.~Taoso and A.~Urbano,
  JCAP {\bf 1412}, no. 12, 045 (2014)
  [arXiv:1407.2173 [hep-ph]].

\bibitem{Ade:2015xua}
  P.~A.~R.~Ade {\it et al.} [Planck Collaboration],
  arXiv:1502.01589 [astro-ph.CO].

\bibitem{Slatyer:2015jla}
  T.~R.~Slatyer,
  arXiv:1506.03811 [hep-ph].

\bibitem{Dutta:2015ysa}
  B.~Dutta, Y.~Gao, T.~Ghosh and L.~E.~Strigari,
  Phys.\ Rev.\ D {\bf 92} (2015) no.7,  075019
  [arXiv:1508.05989 [hep-ph]].


\bibitem{Berlin:2014tja}
  A.~Berlin, D.~Hooper and S.~D.~McDermott,
  Phys.\ Rev.\ D {\bf 89} (2014) no.11,  115022
  [arXiv:1404.0022 [hep-ph]].

\bibitem{Alves:2014yha}
  A.~Alves, S.~Profumo, F.~S.~Queiroz and W.~Shepherd,
  Phys.\ Rev.\ D {\bf 90} (2014) no.11,  115003
  [arXiv:1403.5027 [hep-ph]].

\bibitem{Agrawal:2014una}
  P.~Agrawal, B.~Batell, D.~Hooper and T.~Lin,
  Phys.\ Rev.\ D {\bf 90} (2014) no.6,  063512
  [arXiv:1404.1373 [hep-ph]].

\bibitem{Abdullah:2014lla}
  M.~Abdullah, A.~DiFranzo, A.~Rajaraman, T.~M.~P.~Tait, P.~Tanedo and A.~M.~Wijangco,
  Phys.\ Rev.\ D {\bf 90} (2014) 035004
  [arXiv:1404.6528 [hep-ph]].

\bibitem{Martin:2014sxa}
  A.~Martin, J.~Shelton and J.~Unwin,
  Phys.\ Rev.\ D {\bf 90} (2014) no.10,  103513
  [arXiv:1405.0272 [hep-ph]].

\bibitem{Berlin:2014pya}
  A.~Berlin, P.~Gratia, D.~Hooper and S.~D.~McDermott,
  Phys.\ Rev.\ D {\bf 90} (2014) no.1,  015032
  [arXiv:1405.5204 [hep-ph]].

\bibitem{Basak:2014sza}
  T.~Mondal and T.~Basak,
  Phys.\ Lett.\ B {\bf 744} (2015) 208
  [arXiv:1405.4877 [hep-ph]].

\bibitem{Cline:2014dwa}
  J.~M.~Cline, G.~Dupuis, Z.~Liu and W.~Xue,
  JHEP {\bf 1408} (2014) 131
  [arXiv:1405.7691 [hep-ph]].

\bibitem{Cheung:2014lqa}
  C.~Cheung, M.~Papucci, D.~Sanford, N.~R.~Shah and K.~M.~Zurek,
  Phys.\ Rev.\ D {\bf 90} (2014) no.7,  075011
  [arXiv:1406.6372 [hep-ph]].

\bibitem{Ko:2014loa}
  P.~Ko and Y.~Tang,
  JCAP {\bf 1501} (2015) 023
  [arXiv:1407.5492 [hep-ph]].

\bibitem{Cahill-Rowley:2014ora}
  M.~Cahill-Rowley, J.~Gainer, J.~Hewett and T.~Rizzo,
  JHEP {\bf 1502} (2015) 057
  [arXiv:1409.1573 [hep-ph]].

\bibitem{Freytsis:2014sua}
  M.~Freytsis, D.~J.~Robinson and Y.~Tsai,
  Phys.\ Rev.\ D {\bf 91} (2015) no.3,  035028
  [arXiv:1410.3818 [hep-ph]].

\bibitem{Kaplinghat:2015gha}
  M.~Kaplinghat, T.~Linden and H.~B.~Yu,
  Phys.\ Rev.\ Lett.\  {\bf 114} (2015) no.21,  211303
  [arXiv:1501.03507 [hep-ph]].

\bibitem{Chen:2015nea}
  C.~H.~Chen and T.~Nomura,
  Phys.\ Lett.\ B {\bf 746} (2015) 351
  [arXiv:1501.07413 [hep-ph]].

\bibitem{Gherghetta:2015ysa}
  T.~Gherghetta, B.~von Harling, A.~D.~Medina, M.~A.~Schmidt and T.~Trott,
  Phys.\ Rev.\ D {\bf 91} (2015) 105004
  [arXiv:1502.07173 [hep-ph]].

\bibitem{Cao:2015loa}
  J.~Cao, L.~Shang, P.~Wu, J.~M.~Yang and Y.~Zhang,
  JHEP {\bf 1510}, 030 (2015)
  [arXiv:1506.06471 [hep-ph]].

\bibitem{Freese:2015ysa}
  K.~Freese, A.~Lopez, N.~R.~Shah and B.~Shakya,
  JHEP {\bf 1604} (2016) 059
  [arXiv:1509.05076 [hep-ph]].

\bibitem{Duerr:2015bea}
  M.~Duerr, P.~Fileviez P¨¦rez and J.~Smirnov,
  JHEP {\bf 1606} (2016) 008
  [arXiv:1510.07562 [hep-ph]].

\bibitem{Cai:2015tam}
  Y.~Cai and A.~P.~Spray,
  JHEP {\bf 1606} (2016) 156
  [arXiv:1511.09247 [hep-ph]].

\bibitem{Tang:2015coo}
  Y.~L.~Tang and S.~h.~Zhu,
  JHEP {\bf 1603} (2016) 043
  [arXiv:1512.02899 [hep-ph]].

\bibitem{Ding:2016wbd}
  R.~Ding, Z.~L.~Han, Y.~Liao and W.~P.~Xie,
  JHEP {\bf 1605} (2016) 030
  [arXiv:1601.06355 [hep-ph]].

\bibitem{Krauss:2016cdi}
  M.~E.~Krauss, T.~Opferkuch, F.~Staub and M.~W.~Winkler,
  Phys.\ Dark Univ.\  {\bf 14} (2016) 29
  [arXiv:1605.05327 [hep-ph]].

\bibitem{Escudero:2017yia}
  M.~Escudero, S.~J.~Witte and D.~Hooper,
  arXiv:1709.07002 [hep-ph].


\bibitem{Aartsen:2015zva}
  M.~G.~Aartsen {\it et al.} [IceCube Collaboration],
  arXiv:1510.05223 [astro-ph.HE].

\bibitem{Aartsen:2013bka}
  M.~G.~Aartsen {\it et al.} [IceCube Collaboration],
  Phys.\ Rev.\ Lett.\  {\bf 111} (2013) 021103
  [arXiv:1304.5356 [astro-ph.HE]].

\bibitem{Aartsen:2013jdh}
  M.~G.~Aartsen {\it et al.} [IceCube Collaboration],
  Science {\bf 342} (2013) 1242856
  [arXiv:1311.5238 [astro-ph.HE]].

\bibitem{Aartsen:2014gkd}
  M.~G.~Aartsen {\it et al.} [IceCube Collaboration],
  Phys.\ Rev.\ Lett.\  {\bf 113} (2014) 101101
  [arXiv:1405.5303 [astro-ph.HE]].

\bibitem{Aartsen:2017mau}
  M.~G.~Aartsen {\it et al.} [IceCube Collaboration],
  arXiv:1710.01191 [astro-ph.HE].

\bibitem{Cholis:2012kq}
  I.~Cholis and D.~Hooper,
  JCAP {\bf 1306} (2013) 030
  [arXiv:1211.1974 [astro-ph.HE]].

\bibitem{Anchordoqui:2013dnh}
  L.~A.~Anchordoqui {\it et al.},
  JHEAp {\bf 1-2} (2014) 1
  [arXiv:1312.6587 [astro-ph.HE]].

\bibitem{Murase:2014tsa}
  K.~Murase,
  AIP Conf.\ Proc.\  {\bf 1666} (2015) 040006
  [arXiv:1410.3680 [hep-ph]].

\bibitem{Feldstein:2013kka}
  B.~Feldstein, A.~Kusenko, S.~Matsumoto and T.~T.~Yanagida,
  Phys.\ Rev.\ D {\bf 88} (2013) no.1,  015004
  [arXiv:1303.7320 [hep-ph]].

\bibitem{Esmaili:2013gha}
  A.~Esmaili and P.~D.~Serpico,
  JCAP {\bf 1311} (2013) 054
  [arXiv:1308.1105 [hep-ph]].

\bibitem{Ema:2013nda}
  Y.~Ema, R.~Jinno and T.~Moroi,
  Phys.\ Lett.\ B {\bf 733} (2014) 120
  [arXiv:1312.3501 [hep-ph]].

\bibitem{Bhattacharya:2014vwa}
  A.~Bhattacharya, M.~H.~Reno and I.~Sarcevic,
  JHEP {\bf 1406} (2014) 110
  [arXiv:1403.1862 [hep-ph]].

\bibitem{Higaki:2014dwa}
  T.~Higaki, R.~Kitano and R.~Sato,
  JHEP {\bf 1407} (2014) 044
  [arXiv:1405.0013 [hep-ph]].

\bibitem{Rott:2014kfa}
  C.~Rott, K.~Kohri and S.~C.~Park,
  Phys.\ Rev.\ D {\bf 92} (2015) no.2,  023529
  [arXiv:1408.4575 [hep-ph]].

\bibitem{Esmaili:2014rma}
  A.~Esmaili, S.~K.~Kang and P.~D.~Serpico,
  JCAP {\bf 1412} (2014) no.12,  054
  [arXiv:1410.5979 [hep-ph]].

\bibitem{Fong:2014bsa}
  C.~S.~Fong, H.~Minakata, B.~Panes and R.~Zukanovich Funchal,
  JHEP {\bf 1502} (2015) 189
  [arXiv:1411.5318 [hep-ph]].

\bibitem{Murase:2015gea}
  K.~Murase, R.~Laha, S.~Ando and M.~Ahlers,
  Phys.\ Rev.\ Lett.\  {\bf 115} (2015) no.7,  071301
  [arXiv:1503.04663 [hep-ph]].

\bibitem{Aisati:2015vma}
  C.~El Aisati, M.~Gustafsson and T.~Hambye,
  Phys.\ Rev.\ D {\bf 92} (2015) no.12,  123515
  [arXiv:1506.02657 [hep-ph]].

\bibitem{Boucenna:2015tra}
  S.~M.~Boucenna, M.~Chianese, G.~Mangano, G.~Miele, S.~Morisi, O.~Pisanti and E.~Vitagliano,
  JCAP {\bf 1512} (2015) no.12,  055
  [arXiv:1507.01000 [hep-ph]].

\bibitem{Ko:2015nma}
  P.~Ko and Y.~Tang,
  Phys.\ Lett.\ B {\bf 751} (2015) 81
  [arXiv:1508.02500 [hep-ph]].

\bibitem{Fiorentin:2016avj}
  M.~Re Fiorentin, V.~Niro and N.~Fornengo,
  JHEP {\bf 1611} (2016) 022
  [arXiv:1606.04445 [hep-ph]].

\bibitem{Dev:2016qbd}
  P.~S.~B.~Dev, D.~Kazanas, R.~N.~Mohapatra, V.~L.~Teplitz and Y.~Zhang,
  JCAP {\bf 1608} (2016) no.08,  034
  [arXiv:1606.04517 [hep-ph]].

\bibitem{Chianese:2016smc}
  M.~Chianese and A.~Merle,
  arXiv:1607.05283 [hep-ph].

\bibitem{Cohen:2016uyg}
  T.~Cohen, K.~Murase, N.~L.~Rodd, B.~R.~Safdi and Y.~Soreq,
  arXiv:1612.05638 [hep-ph].

\bibitem{Dhuria:2017ihq}
  M.~Dhuria and V.~Rentala,
  arXiv:1712.07138 [hep-ph].

\bibitem{Bai:2013nga}
  Y.~Bai, R.~Lu and J.~Salvado,
  JHEP {\bf 1601} (2016) 161
  [arXiv:1311.5864 [hep-ph]].

\bibitem{Griest:1989wd}
  K.~Griest and M.~Kamionkowski,
  Phys.\ Rev.\ Lett.\  {\bf 64} (1990) 615.

\bibitem{Hui:2001wy}
  L.~Hui,
  Phys.\ Rev.\ Lett.\  {\bf 86} (2001) 3467
  [astro-ph/0102349].

\bibitem{Doncheski:1997it}
  M.~A.~Doncheski and R.~W.~Robinett,
  Phys.\ Rev.\ D {\bf 56} (1997) 7412
  [hep-ph/9707328].

\bibitem{Anchordoqui:2006wc}
  L.~A.~Anchordoqui, C.~A.~Garcia Canal, H.~Goldberg, D.~Gomez Dumm and F.~Halzen,
  Phys.\ Rev.\ D {\bf 74} (2006) 125021
  [hep-ph/0609214].

\bibitem{Alikhanov:2013fda}
  I.~Alikhanov,
  JHEP {\bf 1307} (2013) 093
  [arXiv:1305.2905 [hep-ph]].

\bibitem{Barger:2013pla}
  V.~Barger and W.~Y.~Keung,
  Phys.\ Lett.\ B {\bf 727} (2013) 190
  [arXiv:1305.6907 [hep-ph]].

\bibitem{Dutta:2015dka}
  B.~Dutta, Y.~Gao, T.~Li, C.~Rott and L.~E.~Strigari,
  Phys.\ Rev.\ D {\bf 91} (2015) 125015
  [arXiv:1505.00028 [hep-ph]].

\bibitem{Dey:2015eaa}
  U.~K.~Dey and S.~Mohanty,
  JHEP {\bf 1604} (2016) 187
  [arXiv:1505.01037 [hep-ph]].

\bibitem{Mileo:2016zeo}
  N.~Mileo, A.~de la Puente and A.~Szynkman,
  JHEP {\bf 1611} (2016) 124
  [arXiv:1608.02529 [hep-ph]].

\bibitem{Dev:2016uxj}
  P.~S.~B.~Dev, D.~K.~Ghosh and W.~Rodejohann,
  Phys.\ Lett.\ B {\bf 762} (2016) 116
  [arXiv:1605.09743 [hep-ph]].

\bibitem{Babu:2001ex}
  K.~S.~Babu and C.~N.~Leung,
  Nucl.\ Phys.\ B {\bf 619}, 667 (2001)
  [hep-ph/0106054].

\bibitem{Rodejohann:2015lca}
  W.~Rodejohann and C.~E.~Yaguna,
  JCAP {\bf 1512}, no. 12, 032 (2015)
  [arXiv:1509.04036 [hep-ph]].

\bibitem{Biswas:2016ewm}
  A.~Biswas, S.~Choubey and S.~Khan,
  JHEP {\bf 1608}, 114 (2016)
  [arXiv:1604.06566 [hep-ph]].

\bibitem{Klasen:2016qux}
  M.~Klasen, F.~Lyonnet and F.~S.~Queiroz,
  arXiv:1607.06468 [hep-ph].

\bibitem{Montero:2007cd}
  J.~C.~Montero and V.~Pleitez,
  Phys.\ Lett.\ B {\bf 675}, 64 (2009)
  [arXiv:0706.0473 [hep-ph]].

\bibitem{Wang:2015saa}
  W.~Wang and Z.~L.~Han,
  Phys.\ Rev.\ D {\bf 92}, 095001 (2015)
  [arXiv:1508.00706 [hep-ph]].

\bibitem{Patra:2016ofq}
  S.~Patra, W.~Rodejohann and C.~E.~Yaguna,
  arXiv:1607.04029 [hep-ph].

\bibitem{Wang:2017mcy}
  W.~Wang, R.~Wang, Z.~L.~Han and J.~Z.~Han,
  Eur.\ Phys.\ J.\ C {\bf 77}, no. 12, 889 (2017)
  [arXiv:1705.00414 [hep-ph]].

\bibitem{Nanda:2017bmi}
  D.~Nanda and D.~Borah,
  arXiv:1709.08417 [hep-ph].

\bibitem{Han:2017ars}
  Z.~L.~Han, W.~Wang and R.~Ding,
  arXiv:1712.05722 [hep-ph].

\bibitem{Arnold:2013cva}
  J.~M.~Arnold, B.~Fornal and M.~B.~Wise,
  Phys.\ Rev.\ D {\bf 88}, 035009 (2013)
  [arXiv:1304.6119 [hep-ph]].

\bibitem{Chang:2016zll}
  W.~F.~Chang, S.~C.~Liou, C.~F.~Wong and F.~Xu,
  JHEP {\bf 1610}, 106 (2016)
  [arXiv:1608.05511 [hep-ph]].

\bibitem{Cacciapaglia:2006pk}
  G.~Cacciapaglia, C.~Csaki, G.~Marandella and A.~Strumia,
  Phys.\ Rev.\ D {\bf 74}, 033011 (2006)
  [hep-ph/0604111].

\bibitem{Aaboud:2017buh}
  M.~Aaboud {\it et al.} [ATLAS Collaboration],
  arXiv:1707.02424 [hep-ex].

\bibitem{ATLAS:2016cyf}
  The ATLAS collaboration [ATLAS Collaboration],
  ATLAS-CONF-2016-045.
  G.~Aad {\it et al.} [ATLAS Collaboration],
  Phys.\ Rev.\ D {\bf 90}, no. 5, 052005 (2014)
  [arXiv:1405.4123 [hep-ex]].

\bibitem{CMS:2016abv}
  CMS Collaboration [CMS Collaboration],
  CMS-PAS-EXO-16-031.
  CMS Collaboration [CMS Collaboration],
  CMS-PAS-EXO-12-061.

\bibitem{Okada:2016gsh}
  N.~Okada and S.~Okada,
  Phys.\ Rev.\ D {\bf 93}, no. 7, 075003 (2016)
  [arXiv:1601.07526 [hep-ph]].

\bibitem{Okada:2016tci}
  N.~Okada and S.~Okada,
  Phys.\ Rev.\ D {\bf 95}, no. 3, 035025 (2017)
  [arXiv:1611.02672 [hep-ph]].

\bibitem{DeRomeri:2017oxa}
  V.~De Romeri, E.~Fernandez-Martinez, J.~Gehrlein, P.~A.~N.~Machado and V.~Niro,
  arXiv:1707.08606 [hep-ph].


\bibitem{Khachatryan:2015vaa}
  V.~Khachatryan {\it et al.} [CMS Collaboration],
  Phys.\ Rev.\ D {\bf 93}, no. 3, 032004 (2016)
  [arXiv:1509.03744 [hep-ex]].
  CMS Collaboration [CMS Collaboration],
  CMS-PAS-EXO-12-041.

\bibitem{CMS:2016qhm}
  CMS Collaboration [CMS Collaboration],
  CMS-PAS-EXO-16-007.
  [CMS Collaboration],
  CMS-PAS-EXO-12-042.

\bibitem{Khachatryan:2016jqo}
  V.~Khachatryan {\it et al.} [CMS Collaboration],
  [arXiv:1612.01190 [hep-ex]].
  A.~M.~Sirunyan {\it et al.} [CMS Collaboration],
  JHEP {\bf 1707}, 121 (2017)
  [arXiv:1703.03995 [hep-ex]].

\bibitem{Aaboud:2016qeg}
  M.~Aaboud {\it et al.} [ATLAS Collaboration],
  New J.\ Phys.\  {\bf 18}, no. 9, 093016 (2016)
  [arXiv:1605.06035 [hep-ex]].

\bibitem{Aad:2011ch}
  G.~Aad {\it et al.} [ATLAS Collaboration],
  Phys.\ Lett.\ B {\bf 709}, 158 (2012)
  Erratum: [Phys.\ Lett.\ B {\bf 711}, 442 (2012)]
  [arXiv:1112.4828 [hep-ex]].

\bibitem{ATLAS:2012aq}
  G.~Aad {\it et al.} [ATLAS Collaboration],
  Eur.\ Phys.\ J.\ C {\bf 72}, 2151 (2012)
  [arXiv:1203.3172 [hep-ex]].

\bibitem{ATLAS:2013oea}
  G.~Aad {\it et al.} [ATLAS Collaboration],
  JHEP {\bf 1306}, 033 (2013)
  [arXiv:1303.0526 [hep-ex]].

\bibitem{Aad:2015caa}
  G.~Aad {\it et al.} [ATLAS Collaboration],
  Eur.\ Phys.\ J.\ C {\bf 76}, no. 1, 5 (2016)
  [arXiv:1508.04735 [hep-ex]].

\bibitem{Khachatryan:2015dcf}
  V.~Khachatryan {\it et al.} [CMS Collaboration],
  Phys.\ Rev.\ Lett.\  {\bf 116}, no. 7, 071801 (2016)
  [arXiv:1512.01224 [hep-ex]].

\bibitem{Aaboud:2017yvp}
  M.~Aaboud {\it et al.} [ATLAS Collaboration],
  Phys.\ Rev.\ D {\bf 96}, no. 5, 052004 (2017)
  [arXiv:1703.09127 [hep-ex]].

\bibitem{Aad:2012tfa}
  G.~Aad {\it et al.} [ATLAS Collaboration],
  Phys.\ Lett.\ B {\bf 716}, 1 (2012)
  [arXiv:1207.7214 [hep-ex]].

\bibitem{Chatrchyan:2012xdj}
  S.~Chatrchyan {\it et al.} [CMS Collaboration],
  Phys.\ Lett.\ B {\bf 716}, 30 (2012)
  [arXiv:1207.7235 [hep-ex]].

\bibitem{Aad:2015zhl}
  G.~Aad {\it et al.} [ATLAS and CMS Collaborations],
  Phys.\ Rev.\ Lett.\  {\bf 114}, 191803 (2015)
  [arXiv:1503.07589 [hep-ex]].

\bibitem{Barger:2006sk}
  V.~Barger, P.~Langacker and G.~Shaughnessy,
  Phys.\ Rev.\ D {\bf 75}, 055013 (2007)
  [hep-ph/0611239].

\bibitem{Barger:2007im}
  V.~Barger, P.~Langacker, M.~McCaskey, M.~J.~Ramsey-Musolf and G.~Shaughnessy,
  Phys.\ Rev.\ D {\bf 77}, 035005 (2008)
  [arXiv:0706.4311 [hep-ph]].


\bibitem{Robens:2015gla}
  T.~Robens and T.~Stefaniak,
  Eur.\ Phys.\ J.\ C {\bf 75}, 104 (2015)
  [arXiv:1501.02234 [hep-ph]].

\bibitem{Robens:2016xkb}
  T.~Robens and T.~Stefaniak,
  Eur.\ Phys.\ J.\ C {\bf 76}, no. 5, 268 (2016)
  [arXiv:1601.07880 [hep-ph]].


\bibitem{Giardino:2013bma}
  P.~P.~Giardino, K.~Kannike, I.~Masina, M.~Raidal and A.~Strumia,
  JHEP {\bf 1405}, 046 (2014)
  [arXiv:1303.3570 [hep-ph]].

\bibitem{Khachatryan:2016vau}
  G.~Aad {\it et al.} [ATLAS and CMS Collaborations],
  JHEP {\bf 1608}, 045 (2016)
  [arXiv:1606.02266 [hep-ex]].

\bibitem{Abdallah:2004wy}
  J.~Abdallah {\it et al.} [DELPHI Collaboration],
  Eur.\ Phys.\ J.\ C {\bf 38}, 1 (2004)
  [hep-ex/0410017].

\bibitem{McDonald:2003zj}
  K.~L.~McDonald and B.~H.~J.~McKellar,
  hep-ph/0309270.

\bibitem{Babu:2002uu}
  K.~S.~Babu and C.~Macesanu,
  Phys.\ Rev.\ D {\bf 67}, 073010 (2003)
  [hep-ph/0212058].

\bibitem{Nebot:2007bc}
  M.~Nebot, J.~F.~Oliver, D.~Palao and A.~Santamaria,
  Phys.\ Rev.\ D {\bf 77}, 093013 (2008)
  [arXiv:0711.0483 [hep-ph]].

\bibitem{Casas:2001sr}
  J.~A.~Casas and A.~Ibarra,
  Nucl.\ Phys.\ B {\bf 618}, 171 (2001)
  [hep-ph/0103065].
  A.~Ibarra and G.~G.~Ross,
  Phys.\ Lett.\ B {\bf 591}, 285 (2004)
  [hep-ph/0312138].

\bibitem{Liao:2009fm}
  Y.~Liao and J.~Y.~Liu,
  Phys.\ Rev.\ D {\bf 81}, 013004 (2010)
  [arXiv:0911.3711 [hep-ph]].

\bibitem{Kohda:2012sr}
  M.~Kohda, H.~Sugiyama and K.~Tsumura,
  Phys.\ Lett.\ B {\bf 718}, 1436 (2013)
  [arXiv:1210.5622 [hep-ph]].

\bibitem{Carpentier:2010ue}
  M.~Carpentier and S.~Davidson,
  Eur.\ Phys.\ J.\ C {\bf 70}, 1071 (2010)
  [arXiv:1008.0280 [hep-ph]].

\bibitem{Nomura:2016ask}
  T.~Nomura and H.~Okada,
  Phys.\ Rev.\ D {\bf 94}, 075021 (2016)
  [arXiv:1607.04952 [hep-ph]].

\bibitem{Guo:2017gxp}
  S.~Y.~Guo, Z.~L.~Han, B.~Li, Y.~Liao and X.~D.~Ma,
  arXiv:1707.00522 [hep-ph].

\bibitem{Dorsner:2016wpm}
  I.~Dorsner, S.~Fajfer, A.~Greljo, J.~F.~Kamenik and N.~Kosnik,
  Phys.\ Rept.\  {\bf 641}, 1 (2016)
  [arXiv:1603.04993 [hep-ph]].

\bibitem{TheMEG:2016wtm}
  A.~M.~Baldini {\it et al.} [MEG Collaboration],
  Eur.\ Phys.\ J.\ C {\bf 76}, no. 8, 434 (2016)
  [arXiv:1605.05081 [hep-ex]].
  J.~Adam {\it et al.} [MEG Collaboration],
  Phys.\ Rev.\ Lett.\  {\bf 110}, 201801 (2013)
  [arXiv:1303.0754 [hep-ex]].

\bibitem{Aubert:2009ag}
  B.~Aubert {\it et al.} [BaBar Collaboration],
  Phys.\ Rev.\ Lett.\  {\bf 104}, 021802 (2010)
  [arXiv:0908.2381 [hep-ex]].

\bibitem{Chakraverty:2001yg}
  D.~Chakraverty, D.~Choudhury and A.~Datta,
  Phys.\ Lett.\ B {\bf 506}, 103 (2001)
  [hep-ph/0102180].

\bibitem{Cheung:2001ip}
  K.~m.~Cheung,
  Phys.\ Rev.\ D {\bf 64}, 033001 (2001)
  [hep-ph/0102238].

\bibitem{Giudice:2012ms}
  G.~F.~Giudice, P.~Paradisi and M.~Passera,
  JHEP {\bf 1211}, 113 (2012)
  [arXiv:1208.6583 [hep-ph]].

\bibitem{Bennett:2006fi}
  G.~W.~Bennett {\it et al.} [Muon g-2 Collaboration],
  Phys.\ Rev.\ D {\bf 73}, 072003 (2006)
  [hep-ex/0602035].

\bibitem{Bauer:2015knc}
  M.~Bauer and M.~Neubert,
  Phys.\ Rev.\ Lett.\  {\bf 116}, no. 14, 141802 (2016)
  [arXiv:1511.01900 [hep-ph]].

\bibitem{Baron:2013eja}
  J.~Baron {\it et al.} [ACME Collaboration],
  Science {\bf 343}, 269 (2014)
  [arXiv:1310.7534 [physics.atom-ph]].

\bibitem{Bona:2007vi}
  M.~Bona {\it et al.} [UTfit Collaboration],
  JHEP {\bf 0803}, 049 (2008)
  [arXiv:0707.0636 [hep-ph]].


\bibitem{feynrules}
  N.~D.~Christensen and C.~Duhr,
  Comput.\ Phys.\ Commun.\  {\bf 180}, 1614 (2009)
  [arXiv:0806.4194 [hep-ph]];
  A.~Alloul, N.~D.~Christensen, C.~Degrande, C.~Duhr and B.~Fuks,
  Comput.\ Phys.\ Commun.\  {\bf 185}, 2250 (2014)
  [arXiv:1310.1921 [hep-ph]].

\bibitem{Belyaev:2012qa}
  A.~Belyaev, N.~D.~Christensen and A.~Pukhov,
  Comput.\ Phys.\ Commun.\  {\bf 184}, 1729 (2013)
  [arXiv:1207.6082 [hep-ph]].

\bibitem{Belanger:2014vza}
  G.~Belanger, F.~Boudjema, A.~Pukhov and A.~Semenov,
  Comput.\ Phys.\ Commun.\  {\bf 192} (2015) 322
  [arXiv:1407.6129 [hep-ph]].

\bibitem{Ade:2013zuv}
  P.~A.~R.~Ade {\it et al.} [Planck Collaboration],
  Astron.\ Astrophys.\  {\bf 571} (2014) A16
  [arXiv:1303.5076 [astro-ph.CO]].

\bibitem{Akerib:2016vxi}
  D.~S.~Akerib {\it et al.} [LUX Collaboration],
  Phys.\ Rev.\ Lett.\  {\bf 118} (2017) no.2,  021303
  [arXiv:1608.07648 [astro-ph.CO]].

\bibitem{Aprile:2017iyp}
  E.~Aprile {\it et al.} [XENON Collaboration],
  Phys.\ Rev.\ Lett.\  {\bf 119}, no. 18, 181301 (2017)
  [arXiv:1705.06655 [astro-ph.CO]].

\bibitem{Cui:2017nnn}
  X.~Cui {\it et al.} [PandaX-II Collaboration],
  Phys.\ Rev.\ Lett.\  {\bf 119}, no. 18, 181302 (2017)
  [arXiv:1708.06917 [astro-ph.CO]].

\bibitem{Cline:2013gha}
  J.~M.~Cline, K.~Kainulainen, P.~Scott and C.~Weniger,
  Phys.\ Rev.\ D {\bf 88}, 055025 (2013)
  Erratum: [Phys.\ Rev.\ D {\bf 92}, no. 3, 039906 (2015)]
  [arXiv:1306.4710 [hep-ph]].

\bibitem{Navarro:1995iw}
  J.~F.~Navarro, C.~S.~Frenk and S.~D.~M.~White,
  Astrophys.\ J.\  {\bf 462} (1996) 563
  [astro-ph/9508025].

\bibitem{Cirelli:2010xx}
  M.~Cirelli {\it et al.},
  JCAP {\bf 1103}, 051 (2011)
  Erratum: [JCAP {\bf 1210}, E01 (2012)]
  [arXiv:1012.4515 [hep-ph]].

\bibitem{Clark:2017fum}
  S.~J.~Clark, B.~Dutta and L.~E.~Strigari,
  arXiv:1709.07410 [astro-ph.HE].

\bibitem{GeringerSameth:2011iw}
  A.~Geringer-Sameth and S.~M.~Koushiappas,
  Phys.\ Rev.\ Lett.\  {\bf 107}, 241303 (2011)
  [arXiv:1108.2914 [astro-ph.CO]].

\bibitem{Ackermann:2011wa}
  M.~Ackermann {\it et al.} [Fermi-LAT Collaboration],
  Phys.\ Rev.\ Lett.\  {\bf 107}, 241302 (2011)
  [arXiv:1108.3546 [astro-ph.HE]].

\bibitem{Ackermann:2013yva}
  M.~Ackermann {\it et al.} [Fermi-LAT Collaboration],
  Phys.\ Rev.\ D {\bf 89}, 042001 (2014)
  [arXiv:1310.0828 [astro-ph.HE]].

\bibitem{Geringer-Sameth:2014qqa}
  A.~Geringer-Sameth, S.~M.~Koushiappas and M.~G.~Walker,
  Phys.\ Rev.\ D {\bf 91}, no. 8, 083535 (2015)
  [arXiv:1410.2242 [astro-ph.CO]].

\bibitem{Ackermann:2015zua}
  M.~Ackermann {\it et al.} [Fermi-LAT Collaboration],
  Phys.\ Rev.\ Lett.\  {\bf 115}, no. 23, 231301 (2015)
  [arXiv:1503.02641 [astro-ph.HE]].

\bibitem{Aguilar:2016vqr}
  M.~Aguilar {\it et al.} [AMS Collaboration],
  Phys.\ Rev.\ Lett.\ {\bf 117}, no. 23,  231102 (2016)


\bibitem{Aartsen:2017sml}
  M.~G.~Aartsen {\it et al.} [IceCube Collaboration],
  arXiv:1701.03731 [astro-ph.HE].

\bibitem{Gandhi:1995tf}
  R.~Gandhi, C.~Quigg, M.~H.~Reno and I.~Sarcevic,
  Astropart.\ Phys.\  {\bf 5} (1996) 81
  [hep-ph/9512364].

\bibitem{Chen:2013dza}
  C.~Y.~Chen, P.~S.~Bhupal Dev and A.~Soni,
  Phys.\ Rev.\ D {\bf 89} (2014) no.3,  033012
  [arXiv:1309.1764 [hep-ph]].

\bibitem{Glashow:1960zz}
  S.~L.~Glashow,
  Phys.\ Rev.\  {\bf 118}, 316 (1960).

\bibitem{Ball:2012cx}
  R.~D.~Ball {\it et al.},
  Nucl.\ Phys.\ B {\bf 867}, 244 (2013)
  [arXiv:1207.1303 [hep-ph]].

\bibitem{Gabriel:1993ai}
  T.~A.~Gabriel, D.~E.~Groom, P.~K.~Job, N.~V.~Mokhov and G.~R.~Stevenson,
  Nucl.\ Instrum.\ Meth.\ A {\bf 338}, 336 (1994).

\bibitem{Kowalski:2004qc}
  M.~P.~Kowalski,
  ``Search for neutrino induced cascades with the AMANDA-II detector,''

\bibitem{Gandhi:1998ri}
  R.~Gandhi, C.~Quigg, M.~H.~Reno and I.~Sarcevic,
  Phys.\ Rev.\ D {\bf 58}, 093009 (1998)
  [hep-ph/9807264].

\bibitem{Ahlers:2015lln}
  M.~Ahlers and F.~Halzen,
  Rept.\ Prog.\ Phys.\  {\bf 78}, no. 12, 126901 (2015).


\bibitem{Aartsen:2015knd}
  M.~G.~Aartsen {\it et al.} [IceCube Collaboration],
  Astrophys.\ J.\  {\bf 809} (2015) no.1,  98
  [arXiv:1507.03991 [astro-ph.HE]].

\bibitem{Aartsen:2014muf}
  M.~G.~Aartsen {\it et al.} [IceCube Collaboration],
  Phys.\ Rev.\ D {\bf 91}, no. 2, 022001 (2015)
  [arXiv:1410.1749 [astro-ph.HE]].


\end{thebibliography}
\end{document}